\documentclass{article}

\usepackage{PRIMEarxiv}

\usepackage[utf8]{inputenc} % allow utf-8 input
\usepackage[T1]{fontenc}    % use 8-bit T1 fonts
\usepackage{hyperref}       % hyperlinks
\usepackage{url}            % simple URL typesetting
\usepackage{booktabs}       % professional-quality tables
\usepackage{amsfonts}       % blackboard math symbols
\usepackage{nicefrac}       % compact symbols for 1/2, etc.
\usepackage{microtype}      % microtypography
\usepackage{lipsum}
\usepackage{fancyhdr}       % header
\usepackage{graphicx}       % graphics
\usepackage{acronym}        % acronyms
\usepackage{siunitx}        % SI units
\usepackage{amsmath}        % formula etc.
\usepackage{algorithm}      % algortihms
\usepackage{algpseudocode}  % algorithms
\usepackage{cleveref}       % easier cross-referencing
\graphicspath{{media/}}     % organize your images and other figures under media/ folder
\usepackage[super,sort&compress,comma]{natbib} 

\title{Multi-fidelity batch Bayesian optimization for bioprocess development across scales}

\author{
    \textbf{Adrian Martens}\\
    Imperial College London\\
    United Kingdom\\
    \And
    \textbf{Mathias Neufang}\\
    Imperial College London\\
    United Kingdom\\
    \And
    Alessandro Butt\'e\\
    DataHow AG\\
    Switzerland\\
    \And
    Moritz von Stosch\\
    DataHow AG\\
    Switzerland\\
    \And
    Antonio del Rio Chanona\\
    Imperial College London\\
    United Kingdom\\
    \And
    Laura M. Helleckes\\
    Imperial College London\\
    United Kingdom\\
    \texttt{l.helleckes@imperial.ac.uk}\\
}

\begin{document}
\maketitle
\begin{abstract}
Bioprocesses are central to modern biotechnology, enabling sustainable production of pharmaceuticals, specialty chemicals, cosmetics, and food. However, developing high-performing processes remains costly and complex, requiring iterative, multi-scale experimentation from microtiter plates to pilot reactors. Conventional Design of Experiments (DoE) approaches often struggle to address process scale-up and the joint optimization of reaction conditions and biocatalyst selection.\\
We present a multi-fidelity batch Bayesian optimization framework to accelerate bioprocess development and reduce experimental costs. The method integrates Gaussian processes tailored for multi-fidelity modeling and mixed-variable optimization. At each iteration, the algorithm proposes not only the next experimental conditions but also the appropriate scale and choice of biocatalyst (i.e., cell clones). To benchmark performance, we developed a custom simulation of a Chinese hamster ovary bioprocess that captures the non-linear, coupled dynamics of scale-up across different clones. Multiple case studies demonstrate that the proposed workflow achieves a reduction in experimental costs while improving yield compared to industrial DoE baselines.
\\
This work provides a data-efficient strategy for bioprocess optimization and highlights opportunities for incorporating transfer learning and uncertainty-aware design for sustainable biotechnology.
\end{abstract}

\section{Introduction}
\label{sec:intro}
Bioprocesses are increasingly used for sustainable manufacturing across various industries, including fine chemicals, food enzymes, and pharmaceuticals~\cite{biodiesel, bioenzymes}.
In addition to replacing established chemical synthesis, bioprocesses provide opportunities for the development of novel pathways and specialty molecules that would be challenging or inefficient to achieve using conventional chemical reactions or other unit operations~\cite{cell_engineering}.
Together with their potential to replace oil-based production routes, these attributes thus give bioprocesses an important role in the drive towards a sustainable bioeconomy in the 21st century.

However, bioprocess development remains a complex and costly undertaking.
Like many processes in the chemical industry, they require a large number of experiments during their development phase, often carried out over multiple stages and scales~\cite{mbr_review1, mbr_review2}.
\subsection{Stages of bioprocess development}
The primary goal of bioprocess development is to efficiently determine the most suitable biocatalysts and the optimal operating conditions for a production process, as well as to successfully transfer them to pilot-scale systems~\cite{high_throughput_cultivation}.
This development pipeline is typically divided into several phases: early- and late-stage screening, optimization, validation, and scale-up.

\begin{figure}[ht!]
    \centering
    \includegraphics[width=0.98\linewidth]{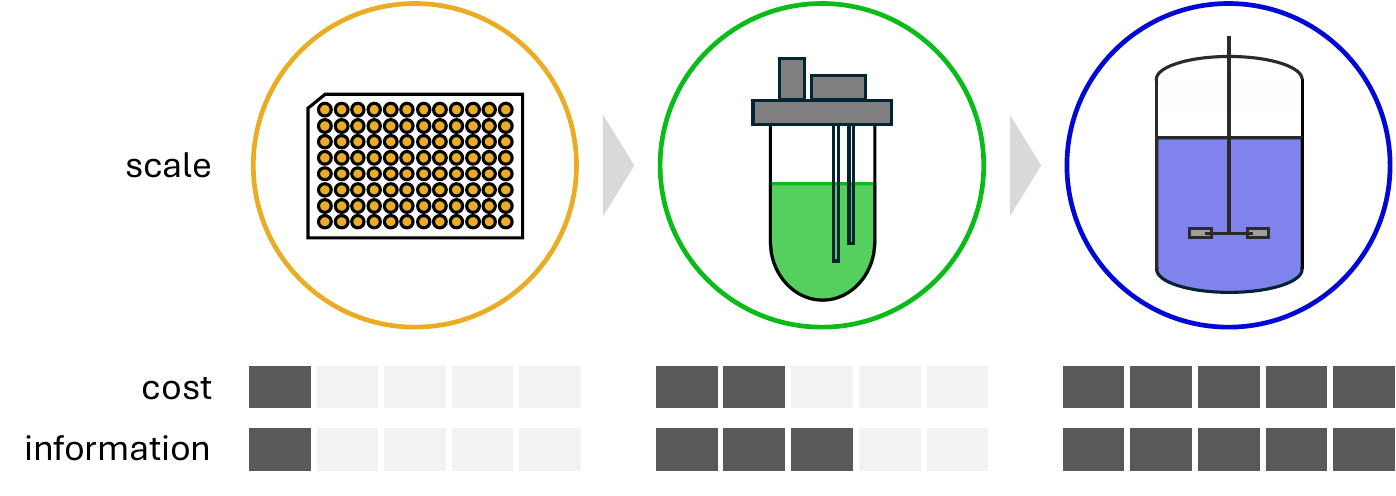}
    \caption{\textbf{Comparison of scales in traditional bioprocess development} \\
    In common primary and secondary screening, cheap and high-throughput experimentation at small-scale (\acp{mtp} in orange and MBRs in green) is achieved at the expense of losing information, i.e., a reduction in available measurements and insights into pilot scale behavior. By weighting cost and information gain, the goal of primary and secondary screening is to maximally reduce the amount of experiments made on the expensive pilot scale (blue).}
    \label{fig:scales}
\end{figure}

Typically, early screening stages are used to reduce the number of potential biocatalyst candidates, i.e.,~the number of strains or cell clones that were previously generated in cell line engineering~\cite{mbr_review3}.
At this stage, experiments are mostly conducted in \acp{mtp}, where operating conditions are either fixed to platform conditions or cannot be controlled at all~\cite{MTP_considerations}.
This stage entails an adequate trade-off between the exploration of suitable candidates and the operating costs, but comes at the expense of limited process control and reduced information about scaled-up behavior~\cite{mbr_review1} (see Figure~\ref{fig:scales}, left).
\\
Following initial screening, the number of clones is reduced to a smaller subset, which is further tested in secondary screening. 
Additional experiments are carried out in shake flasks or \acp{mbr} (see Figure~\ref{fig:scales}, middle). Such high-throughput experimentation systems are available from many suppliers and are used to refine the candidate pool~\cite{ambr}.
These platforms offer greater control over environmental parameters and enable the identification of a lead clone for further development.
The optimization phase, typically still using \ac{mbr}, then focuses on characterizing the selected clone under varying process conditions.
In a final step (see Figure~\ref{fig:scales}, right), pilot plant and scale-down models are in operation to achieve process integration.
In this phase, production capacity is estimated and, in case of pharmaceutical applications, a suitable control strategy is defined~\cite{bioprocess_development_review}.

\subsection{Challenges of traditional process development workflow}
\label{sec:intro_challenges-stage}
The traditional development pipeline follows a funnel-like approach.
As the pipeline progresses, the number of candidate clones and parallel experiments decreases, whereas the level of information per experiment (i.e., the number of available measurements and insights into production behavior) increases.
As the number of controllable process parameters grows with scale, the experimental costs increase simultaneously~\cite{mbr_review1, mbr_review3}.
Along the funnel, each stage is used to narrow down the search space for critical design variables and process parameters~\cite{high_throughput_cultivation}.
However, the influence of these design variables, e.\,g., the cell clone, temperature, pH, formulation, or feeding regime, is often highly non-linear, system-specific, and difficult to quantify due to measurement uncertainty and biological variability~\cite{scale_up_issues_gradients, heterogenity_cell_population}.
In addition, small-scale experiments are generally more homogeneous due to better mixing and thus do not fully account for the heterogeneities present at larger scales.
These challenges result in a large, high-dimensional parameter space with sparse and noisy data, and complicate the transfer of information between scales.

To date, classical \ac{doe} methods are most commonly employed to structure experimental campaigns.
Although these methods are designed to efficiently cover the search space and extract meaningful information for different purposes, most assume very limited interaction between the process parameters~\cite{DoE_review}.
For that reason, most \ac{doe} methods are only useful for their specific use case, with rare cases of more advanced applications in optimization problems.
However, bioprocesses are highly complex and typically characterized by various non-linear relationships and parameter interdependencies, thus requiring advanced process models and intricate experimental designs for sufficient characterization and transfer of knowledge between scales~\cite{helleckes_bayesian_2022}.

Addressing these challenges requires optimization strategies that are sample-efficient, robust to uncertainty, and propagate information across scales.
Traditional heuristic approaches or \ac{doe} techniques, such as the dominant methods of Box-Behnken, Definitive Screening or Full Factorial Designs for Response Surface Models~\cite{DoE_review}, are limited in their ability to discover optimal process parameter settings under such constraints~\cite{DoE_review}.
\ac{bo} offers a compelling alternative as a data-efficient, model-based approach.
\ac{bo} uses probabilistic surrogate models and acquisition functions to balance exploration and exploitation in the search space~\cite{BO_overview, siska_guide_2025}.
Its ability to incorporate uncertainty and adaptively guide experimentation has made it a valuable tool in domains with expensive and noisy evaluations, including hyperparameter tuning for machine learning models~\cite{hyper_pram_tuning}, design of control systems~\cite{controls_BO}, and chemical engineering applications such as reactor design~\cite{reactor_design} and experiment planning~\cite{experimentation_planning}.
Most recently, \ac{bo} also gained traction in bioprocess engineering~\cite{bo_in_bioprocesses, thompson_integrating_2023}.
Within this field, advanced \ac{bo} techniques such as batch and multi-fidelity allow to model uncertainty and optimize across different scales in a holistic approach, rather than limiting the propagation of information to a sequential approach as currently done in industry.

\subsection{Aim of this study}
This paper introduces a systematic approach to multi-fidelity batch \ac{bo}, specifically adapted to bioprocess development.
First, several case studies inspired by real-world industrial conditions are introduced, which are used to compare the \ac{bo} approaches to the traditional industrial approach of \ac{doe}.
Second, different acquisition functions for \ac{bo} across scales are tested, highlighting the potential over the traditional, funnel-like approach.
In detail, \ac{bo} accounts for the stepwise nature of process scale-up, incorporates multiple fidelity levels corresponding to different scales of parallel experimentation, and handles mixed-variable input spaces, i.e.,~the optimization of both biocatalysts and process conditions.
The benchmarking uses a simulation-based model of a \ac{cho} cell process, a workhorse in biopharmaceutical applications~\cite{CHO_review, CHO_review_2, jimenez_del_val_chompact_2023}.
The generated data is used to demonstrate how the \ac{bo} workflow can improve the efficiency of the bioprocess development pipeline compared to standard industrial approaches.

\section{Methods}
\label{sec:methods}
\subsection{Bioprocess simulation models}
\label{sec:bioprocess_model}
\subsubsection{Base Model}
The core of the bioprocess model is a system of \acfp{ode} with model parameters empirically identified in \cite{bio_system_base}. It captures the dynamic growth behavior of \ac{cho} cells and their associated substrate kinetics, including the consumption and production of glucose, glutamine, ammonia, and lactate, using Monod-type kinetic equations. By treating the system as an initial value problem, it is possible to compute the complete concentration trajectories of the total biomass, as well as the individual substrates over time. However, modifications to the original model are necessary, as the base formulation assumes standard conditions and fixed process parameter values. In contrast, these are key design variables in real industrial scenarios and must be treated as tunable inputs to enable meaningful optimization and process control.

\subsubsection{Product Modeling}

Since the base model focuses solely on cell dynamics, it is necessary to introduce product formation terms to capture the actual performance of the bioprocess. Inspired by the formulation of product generation in Pennington \textit{et al.}~\cite{bioprocess_product}, additional parameters and an equation describing the production of an arbitrary product molecule are incorporated into the bioprocess simulation model:
\begin{equation}
    \frac{dC_\text{P}}{dt} = Y_{\text{P,X}}C_\text{X,V} + Y'_{\text{P,X}}\left({\mu_\text{max}} \frac{C_\text{G}}{K_\text{G} + C_\text{G}} \right) C_\text{X,V},
\end{equation}
where $C_\text{P}$ denotes the product concentration, $C_\text{X,V}$ the viable biomass concentration, $C_\text{G}$ the glucose concentration, $K_{\text{G}}$ the half velocity of the product, and $\mu_\text{max}$ the maximum growth rate. The parameters $Y'_{\text{P,X}}$ and $Y_{\text{P,X}}$ represent growth-associated and non-growth-associated product yields, respectively. This formulation is seamlessly integrated into the existing system of \acp{ode} governing the model dynamics.

\subsubsection{Temperature and pH Influence Modeling}

Temperature and pH significantly affect cell growth through mechanisms more complex than those of substrates. Their influence can lead to strong inhibition or prevent growth entirely if chosen to be outside of the tolerable range for a specific cell. Furthermore, the temperature impact can be asymmetric and non-monotonic. While temperatures that are too low can merely slow down cellular activities, temperatures above the cell's specific tolerance threshold can lead to abrupt drops in activity or even protein denaturation \cite{price_temperature_2004, temperature_dependence}.

This behavior was modeled as a bell-shaped function for the influence factor:
\begin{equation}  
    k(x) = A \cdot 
    \begin{cases}
    \exp\left(-L_1 \left(\frac{x - x_\text{opt}}{L_2}\right)^2\right), & \text{if } x < x_\text{opt} \\
    \exp\left(-R_1 \left(\frac{x - x_\text{opt}}{R_2}\right)^2\right), & \text{if } x \geq x_\text{opt},
    \end{cases}
\end{equation}
which is multiplied by the model's Monod maximum growth rate \(\mu_{max}\), yielding the condition-corrected growth rate:
\begin{align*}
    \hat{\mu}(T, pH) = \mu_{\text{max}} \cdot k_{\text{temperature}}(T) \cdot k_{\text{pH}}(pH).
\end{align*}
The shape of the influence parameter function is determined by the positive parameters $L_1,L_2,R_1$ and $R_2$. The $L$ parameters determine the steepness and smoothness of the drop-off of the left side, and $R$ do the same for the right side of the function. $A$ is a scaling factor that sets the value at the maximum, while $x_{opt}$ determines the position of the maximum.

\begin{figure}[ht!]
    \centering
    \includegraphics[width=1\linewidth]{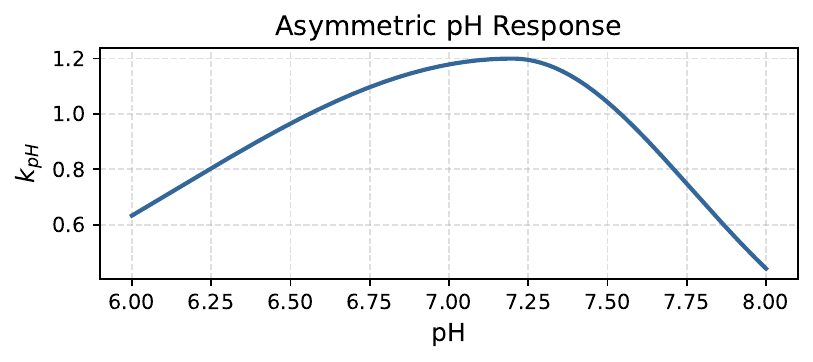}
    \caption{\textbf{Influence factor behavior over pH range} \\
    This plot depicts the behavior of pH influence factor $k_{\text{pH}}(pH)$. The factor affects the maximum growth rate $\mu_\text{max}$ for a specific cell clone type. Depending on the pH value of the reaction conditions, cell growth is inhibited ($k_{\text{pH}}(pH) < 1$) or promoted ($k_{\text{pH}}(pH) > 1$). This curve is created through an exemplary parameter combination for a specific clone: $x_{\text{opt}}=7.2$, $L_1=1$, $L_2=1.5$, $R_1=1$, $R_2=0.8$ and $A=1.2$}
    \label{fig:pH_influence}
\end{figure}

The parameters for the correction factor functions (see \Cref{tab:influence_factors}) were chosen such that the temperature influence follows a smoother distribution than that of pH, since cells usually have much tighter constraints on pH. The exemplary result for the pH factor curve can be seen in \Cref{fig:pH_influence}.

\begin{table}[ht!]
  \centering
  \small
  \caption{\textbf{Description and values of parameters for pH and temperature influence}}
  \label{tab:influence_factors}
  \begin{tabular*}{0.62\textwidth}{@{\extracolsep{\fill}}lll}
    \hline
    Variable & Description & Value \\
    \hline
    $x$ & Actual pH or T value & input \\
    $x_{\text{opt}}$ & Optimal pH or T value & clone-dependent \\
    $A$ & Scaling constant & 1.2 \\
    $(L_1, L_2)$ & Decay parameters for $x < x_\text{opt}$ & $(1,5)$ for T; $(1,1.5)$ for pH \\
    $(R_1, R_2)$ & Decay parameters for $x \geq x_\text{opt}$ & $(1,5)$ for T; $(1,0.8)$ for pH \\
    \hline
  \end{tabular*}
\end{table}

\subsubsection{Clone Modeling}

To simulate clone diversity, clones are generated by randomly sampling model parameters oriented around base values taken from \cite{bio_system_base, bioprocess_product} (see \Cref{sec:clone_parameters}).30 clones are chosen to provide sufficient behavioral variability in the process model while keeping the computational cost for surrogate models within a feasible range. Any variations introduced mimic the differences in productivity, robustness, and growth characteristics typically observed in real clone libraries. \Cref{sec:results_and_discussion} further investigates how the underlying distribution of clone performance affects the optimization outcomes.

\subsubsection{Feed Modeling}

In the simulation study, a fed-batch reactor configuration is employed, where experiments receive a total of three feeding pulses at fixed points in time. However, a distinction is made in the feeding strategy based on the reactor type: microtiter plate (\ac{mtp}) experiments are limited to a single feeding pulse varying in amount at the second point in time, whereas the more flexible microbioreactor (\ac{mbr}) and pilot-scale reactors are capable of accommodating all three pulses. These pulses are administered as bolus feeds at fixed time points and are considered instantaneous. The amount of substrate added per pulse is denoted as $F_1$, $F_2$, and $F_3$, respectively, and the total available substrate $\sum_{i=1}^{3}F_i$ is limited by a maximum allowable feed amount $F_{\text{max}}$. Furthermore, ideal mixing, an immediate concentration change, and no change in reactor volume upon feeding are assumed. These simplifications allow the model to remain simple while still capturing the impact of different feeding strategies. 

The implementation was carried out using piecewise numerical integration up to each respective feeding time point. At each pulse, the relevant substrate concentrations are increased by the corresponding feed amount, after which the system is re-integrated from the updated state until the next feeding event.

\subsubsection{Scale Modeling} \label{sec:scale_modeling}
It is assumed that three experimentation scales are available: \acf{mtp} ($<$\SI{1}{\milli\liter}), \acf{mbr} (inspired by the ambr250~\cite{bareither_automated_2013}, a \SI{250}{\milli\liter} stirred reactor), and pilot scale (inspired by a \SI{20}{\liter} reactor). To reflect scale-dependent influences on process behavior, two additional parameters $s_{\text{growth}}$ and $s_{\text{death}}$ are introduced. The first scale parameter, $s_{\text{growth}}$, represents growth-inhibiting effects and captures deviations from the ideal growth conditions typically observed across reactor scales. It is multiplied by the maximum growth rate $\mu_{\text{max}}$, in the same way as the temperature and pH correction factors $k_{\text{temperature}}$ and $k_{\text{pH}}$. For instance, even if environmental factors promote growth, $s_{\text{growth}}$ can counterbalance such positive effects to reflect limitations specific to a given scale:
\begin{equation}
    \hat{\mu}(T, pH) = \mu_{\text{max}} \cdot k_{\text{temperature}}(T) \cdot k_{\text{pH}}(pH) \cdot s_{\text{growth}}
\end{equation}

To account for scale-dependent differences in cell death behavior, the second scale parameter, $s_{\text{death}}$, is introduced and applied to the base model expression for cell death rate $k_{\text{d}}$. This allows for an increase in the death rate from the base case, depending on the experimental scale:
\begin{equation}
    k_{\text{d}} = k_{\text{d,max}} \cdot \left(\frac{k_\mu}{\hat{\mu}(T, pH) + k_\mu}\right) \cdot s_{\text{death}},
\end{equation}
where $k_{\text{d}}$ is the effective death rate, $k_{\text{d,max}}$ is the maximum death rate, and $k_\mu$ is a model parameter reflecting intrinsic death dynamics. Note that the corrected growth rate $\hat{\mu}$ also influences the death term.

Both of these effects come together in the viable cell differential equation, where $X_V$ is the time-dependent viable cell density in the system:
\begin{equation}
    \frac{dX_V}{dt}=(\hat{\mu}-k_d) \cdot (X_V)
\end{equation}

For the pilot scale, $s_{\text{growth}}$ and $s_{\text{death}}$ are assumed to reflect the effects of process inhomogeneities (such as gradients in substrate or oxygen concentration) and higher shear stress caused by larger-scale mixing. These scale-related challenges are more prevalent in pilot reactors and are collectively captured by the scaling factors \cite{heterogenity_cell_population, scale_up_issues_gradients}.

\ac{mtp} experiments are similarly affected by scale-related limitations. Here, the scaling parameters aim to capture the negative impact, for example due to error-prone manual preparation of \acp{mtp} and variability in the preparation of the cell clones. Moreover, the small volumes used in \ac{mtp} batches lead to a higher influence of pipetting errors and limit the ability to accurately measure and control the process.

\ac{mbr} experiments are considered to offer a good compromise between system size and experimental control, since they are miniaturized stirred reactors. It is assumed that they are less affected by both initialization issues (as in \ac{mtp}) and large-scale inhomogeneities (as in the pilot), and thus no significant growth-inhibiting scale effects are introduced for this scale.

In addition to the scaling factors, measurement noise is introduced in the model as reactor-scale-dependent additive Gaussian noise to account for reduced precision in evaluating the cumulative product titer. This effect is most pronounced for the smallest scale (\ac{mtp}), due to technical limitations in accurately measuring concentrations in very small volumes. In contrast, measurement inaccuracies are assumed to be significantly lower at the \ac{mbr} and pilot scales.

After carefully taking into account all scale-dependent effects, values for $s_{\text{growth}}$, $s_{\text{death}}$ and measurement noise were chosen, which are listed in \Cref{tab:inhibition_factors}. With that, the \ac{mbr} scale is modeled to resemble a scaled-down version of the pilot reactor.

\begin{table}[h]
  \centering
  \caption{\textbf{Chosen values for the scale-related modeling parameters and the measurement noise}\\
  Values for $s_{\text{growth}}$ and $s_{\text{death}}$ were chosen such that the \ac{mtp} scale is most unreliable, followed by the pilot scale, and the \ac{mbr} scale with best performance.}
  \label{tab:inhibition_factors}
  \begin{tabular*}{0.48\textwidth}{@{\extracolsep{\fill}}cccc}
    \hline
    Scale & $s_{\text{growth}}$ & $s_{\text{death}}$ & Measurement noise scale \\
    \hline
    \ac{mtp} & 0.5 & 1.1 &  $1.5 \cdot 10^{-1}$\\
    \ac{mbr} & 1.0 & 1.0 & $5 \cdot 10^{-3}$ \\
    Pilot    & 0.9 & 1.3 & $5 \cdot 10^{-3}$ \\
    \hline
  \end{tabular*}
\end{table}

\subsubsection{Cost Modeling}
In addition to the different effects on bioprocess performance, the costs associated with experiments at different scales also vary significantly. Following the cost estimation in \ref{sec:cost_estimation}, the costs per experiment are assumed to be approximately €10, €500, and €2000 for \ac{mtp}, \ac{mbr}, and pilot-scale reactors, respectively. These values are not intended as exact cost determination, but rather as indicative approximations to reflect the order of magnitude of scale-dependent cost differences.

\subsection{Bayesian Optimization for Bioprocess Development}
\subsubsection{Bayesian Optimization}
Optimizing black-box objective functions using only noisy zeroth-order information can be challenging. For this purpose, many \ac{ddo} methods have been developed, which effectively tackle the problem through approaches such as surrogate models or numerical derivative approximations \cite{DDO_review}. However, most methods require numerous evaluations of the objective function or rely on additional knowledge about the problem. As a result, in scenarios involving expensive evaluations, computational cost becomes the primary bottleneck, and sample efficiency is critical.

\ac{bo} is an optimization framework that enables sample-efficient optimization of objective functions that are costly to evaluate. A central idea in \ac{bo} is the use of a probabilistic surrogate model that expresses a belief about the objective function. This belief is formalized through a prior distribution over functions, which is updated as observations are collected, resulting in a posterior distribution. The posterior captures the updated belief about the objective function's behavior after accounting for the observed data. The posterior predictive distribution, which describes the predictive uncertainty of the surrogate model in the search space, plays a central role in decision making. It is used by \ac{bo}'s acquisition function to trade off between exploration (sampling where uncertainty is high) and exploitation (sampling where expected performance is high). An acquisition function in this context is a heuristic or decision rule that guides the optimization process. The next evaluation point is selected as the maximizer of the acquisition function, and the objective function is then evaluated at that location. This cycle of belief updating and point selection is repeated until a resource limit (typically a maximum number of iterations or total experimental cost) is reached. Otherwise, if sufficient information about the optimum is gained, early termination is possible. Further fundamental information on \ac{bo} can be found in \cite{garnett_bayesoptbook_2023, BO_overview}.

The \ac{bo} framework, often known in the form of the Efficient Global Optimization algorithm \cite{classic_EI}, has undergone numerous extensions. These include advances in acquisition functions, evaluation strategies, and surrogate models \cite{KG_example, review_BO_advancements}. Choosing the right model type is a critical aspect for successful optimization.

\subsubsection{Gaussian Processes}
Among the available options, \acp{gp} are the most commonly used surrogate models in \ac{bo} due to their closed-form expressions, flexibility, interpretability, and ability to provide uncertainty estimates alongside predictions \cite{Rassmussen_Williams}. As non-parametric regression models, \acp{gp} adapt their complexity to the amount of available data, making them effective in scenarios with limited observations. In addition, they provide principled predictive uncertainty, which is essential in \ac{bo}. For the following formal introduction a notation similar to the one of Rasmussen et al.~is used \cite{Rassmussen_Williams}.

A \ac{gp} defines a distribution over functions:
\begin{equation}
    f(\mathbf{x}) \sim \mathcal{GP}(m(\mathbf{x}), k(\mathbf{x}, \mathbf{x}')),
\end{equation}
where $m(\mathbf{x}):\mathbb{R}^d \to\mathbb{R}$ with $\mathbf{x}\in\mathbb{R}^d$ is the mean function and $k(\mathbf{x}, \mathbf{x}'):\mathbb{R}^d \times \mathbb{R}^d \to \mathbb{R}$ is the covariance (or kernel) function. This implies that for any finite set of input points $\{\mathbf{x}_1, \ldots, \mathbf{x}_n\}$, the corresponding outputs $\mathbf{f} = \{f(\mathbf{x}_1), \ldots, f(\mathbf{x}_n)\}$ follow a multivariate normal distribution:
\begin{equation}
    \mathbf{f} \sim \mathcal{N}(\boldsymbol{m}, \mathbf{K}),
\end{equation}
with $m_i = m(\mathbf{x}_i)$ and $K_{ij} = k(\mathbf{x}_i, \mathbf{x}_j)$.
The kernel function $k(\mathbf{x}, \mathbf{x}')$ encodes the similarity between inputs and thus governs the smoothness and generalization behavior of the \ac{gp}. Most commonly, the kernel depends on the Euclidean distance between input points. However, a variety of kernel functions exists, and kernel design can be seen as a discipline of its own \cite{tanimoto_kernel, deep_kernel_learning}. Ultimately, a \ac{gp} must be tailored to the specific characteristics of the problem it is modeling to perform effectively.

\subsubsection{Multi-Fidelity}
In addition to modeling the influence of key variables, it is equally important to account for the varying quality and resolution of the available data. Multi-fidelity modeling captures these inherent differences in scale, evaluation cost, precision, and information gain that arise in real-world experimental and simulation scenarios. It can be seen as a form of transfer learning, where information acquired from one task or domain is transferred to enhance learning in a related task \cite{transfer_learning}. This concept is particularly applicable to experiments conducted at different physical or conceptual scales, where simplified or idealized setups are used to infer properties of the target system.

To collect data economically, low-cost experiments or simulations are often employed, which typically involve idealizations or miniaturization of the actual target system. These low-fidelity sources may be fast or cheap to evaluate, but yield only approximate information about the objective function. Contrary, high-fidelity sources are more costly but offer more accurate and reliable information.
In the funnel-approach during bioprocess development, \ac{mtp}, \ac{mbr} and pilot-scale experiment exactly reflect such a case of different fidelities (compare \Cref{fig:scales}).

There are several strategies for incorporating information from multiple fidelity levels into a unified model \cite{review_multi_fidelity}. These include methods such as multi-task models \cite{multi_task_1, multi_task_2}, hierarchical Kriging \cite{hirarchical_kringing}, acquisition-function-based fidelity selection \cite{multi_fidelity_acquisition_function}, and fidelity modeling as an explicit input variable. The focus of this work is on the latter: the fidelity level is modeled as an additional input dimension in the \ac{gp}, thereby embedding fidelity into the structure of the search space.

Fidelity levels can be treated as continuous or discrete variables. Continuous fidelity levels are represented e.g. in the amount of data for cross-validation of neural network performance or the number of training iterations \cite{continuous_fidelity}. Discrete fidelity levels can be used to capture differing reactor scales and are thus employed in this work. These discrete fidelity levels can be ordered based on their expected information quality regarding the objective function, and a qualitative measure of their similarity can also be assumed. Under these assumptions, discrete fidelity levels can be mapped onto a continuous fidelity axis. Conversely, a continuous fidelity spectrum could be discretized into discrete fidelity levels. The relative spacing between levels reflects the similarity in system behavior rather than the absolute fidelity quality. Therefore, the actual numeric values assigned to fidelity levels serve to encode a learnable structure via the \ac{gp} kernel’s lengthscales. 

\begin{figure}[ht!]
    \centering
    \includegraphics[width=0.8\linewidth]{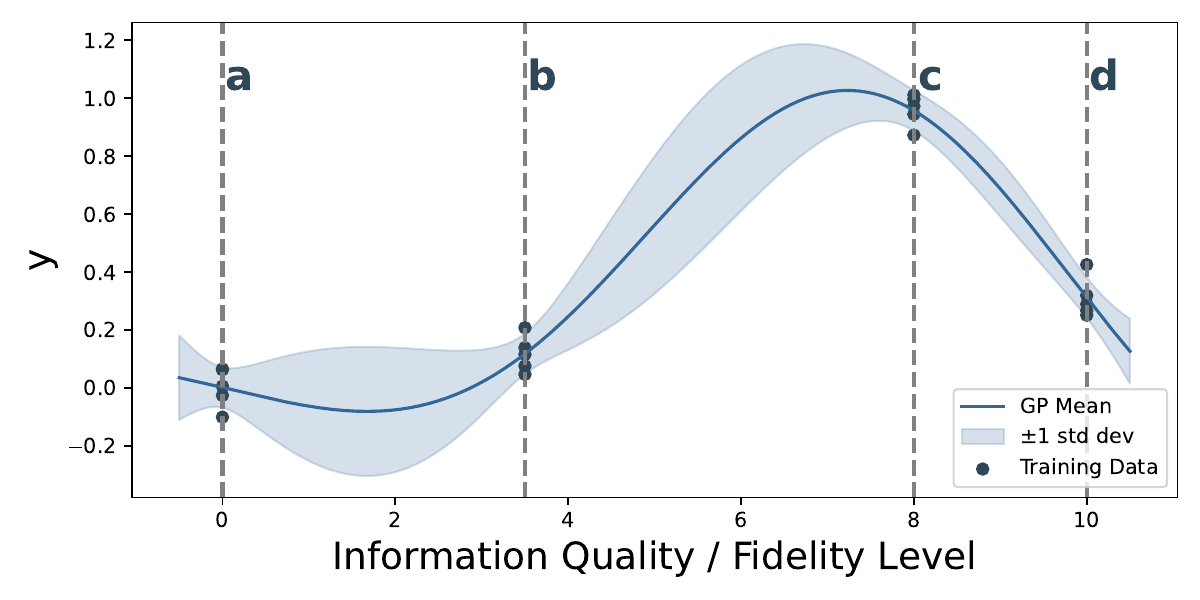}
    \caption{\textbf{Exemplary dimension axis "Information Quality"  with discrete fidelity levels fixed at specific values} \\
    The plot depicts a possible distribution of the fidelity levels a-d with varying quality of information. Each fidelity level has five measurements and collectively a \ac{gp} was fitted to this data. The proximity of fidelity levels determines the degree of cross-covariance, such that observations at one fidelity level exert greater influence on nearby levels. This reflects the model’s assumption of correlated behavior across fidelity levels based on their relative similarity. Consequently, the distance between two levels can be changed to also change the strength of their interaction.}
    \label{fig:multi_fidelity_example}
\end{figure}

In this study, the fidelity axis is assumed to be scalable and learnable in a practical range (here $[0, 10]$, which is standardized in preprocessing for the \ac{gp}). An example of this modeling approach is shown in \Cref{fig:multi_fidelity_example}, which illustrates four discrete fidelity levels $a$ through $d$, ordered such that $d > c > b > a$ in terms of information quality. Important for this approach is the match between the fidelity model and the actual differences in the system. A key requirement for this approach is that the fidelity dimension must realistically reflect the actual differences between data sources. Only then can the \ac{gp} provide reliable predictions and guide the acquisition function effectively. Since the \ac{mbr} scale is assumed to be a relatively accurate scale-down model of the pilot system, we position it closer to the pilot in the \ac{gp}’s fidelity space (at 7 between 0 and 10).

\subsubsection{Batch Evaluation}
\label{sec:batch_modeling}
Beside different scales of experimentation, another aspect in bioprocess development is the parallelization of experiments. Batch \ac{bo} refers to the selection of multiple points for evaluation at the same time instead of a sequential selection of one point at a time. In the context of bioprocess optimization, batch-wise planning makes the experimentation more efficient and leverages the technologies and methods available for high-throughput experimentation \cite{mbr_review2, high_throughput_exp_example, gonzalez_new_2023}.

However, this efficiency gain comes at the cost of more complex batch selection. This is because, unlike in the sequential case, candidate points within a batch are not conditionally independent: the predictions made by the \ac{gp} model for each point are correlated due to the shared posterior covariance structure. As a result, evaluating one point in the batch affects the expected utility of evaluating others. These interdependencies mean that most useful acquisition functions cannot select each point in a batch independently.

The batch size is commonly denoted as $q$, which means that iterative evaluations of batches of size $q=1$ reflect the sequential case. Since for the acquisition function this implies finding the one-step optimal point, most of the time analytical expressions for calculating acquisition function values are available. Unfortunately, for the general case of batch size $q>1$, the calculations can often only be approximated with more advanced sampling or penalization methods and depend on the specific requirements of each acquisition function.

Fantasizing, also known as the Kriging Believer or Constant Liar approach \cite{kringing_believer}, is a common strategy for optimizing multi-point acquisition functions. It enables efficient batch selection by approximating the joint acquisition process through a sequential heuristic. Initially, the acquisition function is optimized to select a single point (i.e., the sequential $q = 1$ case). The \ac{gp} model is then temporarily conditioned on a "fantasized" observation at that point (typically based on a sample from the posterior). Using this updated model, a second point is selected, and the process is repeated until a full batch of $q$ points is generated. To improve robustness, this procedure is performed over multiple fantasy samples, and the acquisition values are averaged. While this greedy approximation avoids the computational burden of evaluating the joint acquisition over all batch points, its performance can degrade with increasing batch size due to the accumulation of approximation error.

Another method used for optimizing multi-point acquisition functions in this work is Monte Carlo sampling. In this approach, multiple function samples are drawn from the joint posterior distribution of the surrogate model, and the acquisition function value of the batch is approximated by averaging over these samples \cite{balandat_botorch_2020}.

These approaches are implemented in \texttt{qUpperConfidenceBound}, \texttt{qLogExpectedImprovement}, \texttt{qMultiFidelityLowerBoundMaxValueEntropy} and BoTorch optimization methods. We advise the interested reader to see implementation details in the respective documentations.

\subsubsection{Categorical Variables}
\label{sec:categorical_variables}
The modeling approaches presented in this section are again inspired by principles of transfer learning. In particular, different cell clones are treated as distinct but related tasks, which can be modeled using multi-task \ac{gp} kernels \cite{multi_task_1}. Alternatively, a simple learned embedding can be used, which allows to calculate Euclidean distances between clones \cite{entity_embedding}.

A commonly used kernel for multi-task \acp{gp} is the \ac{icm} kernel \cite{multi_task_1}. The \ac{icm} kernel learns a low-rank representation of the correlations between tasks by combining a shared base kernel with a learned task covariance structure. In this context, the base kernel encodes the similarity in the input space, while the coregionalization matrix captures how strongly the behavior of one clone relates to another.

Mathematically, the \ac{icm} kernel is defined as:

\begin{equation}
k((\mathbf{x}, t), (\mathbf{x}', t')) = k_\text{base}(\mathbf{x}, \mathbf{x}') \cdot \mathbf{B}_{t, t'}
\end{equation}

Here, $\mathbf{x}, \mathbf{x}'$ denote input parameter vectors (e.\,g., pH, temperature, feed rate), $t, t'$ denote the categorical task indices (in this case, the cell clone type), $k_\text{base}$ is a standard kernel such as the \ac{rbf} kernel, and $\mathbf{B} \in \mathbb{R}^{T \times T}$, $T$ being the number of tasks, is the positive semi-definite coregionalization matrix that is learned from the data during training.

This multiplicative combination of a base kernel with a clone-specific covariance structure allows the model to capture the global input-dependent behavior as well as clone-specific deviations from it. This lets the \ac{gp} transfer information across clones while accounting for their individual characteristics, effectively leveraging similarities between them to improve data efficiency and prediction accuracy during optimization.

Another approach is using an embedding, which does not directly learn a covariance structure but instead learns positions in an $n$-dimensional space. The distances between these positions are then used to compute a covariance-like factor. Each categorical variable (e.\,g., cell clone) is represented as a one-hot vector, which is mapped to a continuous embedding via a learned matrix. This setup allows for convenient computation of squared distances, which are then transformed using a squared exponential-like kernel. The overall kernel function combines this categorical similarity with the base kernel for continuous input features:

\begin{equation}
k((\mathbf{x}, t), (\mathbf{x}', t')) = k_\text{base}(\mathbf{x}, \mathbf{x}') \cdot \exp\left(-\frac{1}{2} \left\lVert \mathbf{E}\mathbf{e}_t - \mathbf{E}\mathbf{e}_{t'} \right\rVert^2\right)
\end{equation}

Here, $\mathbf{e}_t$ and $\mathbf{e}_{t'}$ are the one-hot encodings of the categorical values $t$ and $t'$, respectively, and $\mathbf{E} \in \mathbb{R}^{T \times T}$ is the learnable embedding matrix. The parameters of $\mathbf{E}$ are optimized jointly with the other hyperparameters of the \ac{gp} by maximizing the model's marginal log-likelihood.

Both the \ac{icm} kernel as well as the representation learning via an embedding matrix $\mathbf{E}$ are compared in case studies in \ref{sec:results_and_discussion} in more detail. With these modifications, the intricate aspects of the biosystem are reflected in the surrogate model, which therefore can provide better estimates of the acquisition function in the \ac{bo} framework.

\subsection{Acquisition functions}
\label{sec:acquisition_functions}
Acquisition functions define the decision rule by which candidate points are selected for evaluation in each iteration of \ac{bo}. These functions leverage both the predictive mean and the covariance (uncertainty) provided by the \ac{gp} model to quantify the potential benefit of sampling at different locations in the search space. Typically, the next evaluation point is chosen by maximizing the acquisition function, thereby balancing the trade-off between exploring uncertain regions and exploiting areas with high predicted performance. 

The following section provides a brief overview of batch Expected Improvement \cite{LogEI_Meta}, batch Upper Confidence Bound \cite{qUCB} and their multi-fidelity extensions, as well as \ac{gibbon} \cite{GIBBON}. These are the acquisition functions employed in this work. For detailed mathematical derivations and theoretical foundations, the reader is referred to the respective literature.

\subsubsection{Extended Multi-Point Log Expected Improvement (extended qLogEI)}
The \ac{ei} criterion originally proposed in \cite{EI_origins} is one of the most widely used acquisition functions in \ac{bo} and has a closed-form expression in the "classic" single-point evaluation ($q=1$) case \cite{classic_EI}:
\begin{equation}
    \text{EI}_{y^*}(\mathbf{x}|\mathcal{D})= \mathbb{E}_{f(\mathbf{x})}\left[\left[f(\mathbf{x})-y^*\right]_+\right]=\sigma(\mathbf{x},\mathbf{x})\left[z\Phi(z) + \phi(z)\right]
\end{equation}
with $z=\left(\frac{\mu(\mathbf{x})-y^*}{\sigma(\mathbf{x},\mathbf{x})}\right)$ where $y^*$ is the incumbent and $[\cdot]_+$ is $\max(0,\cdot)$. $\mathcal{D} = \left\{ (\mathbf{x}_i, y_i) \right\}_{i=1}^{N}$ denotes the set of the $N$ already observed points, with $\mathbf{x} \in \mathbb{R}^d$ being any new input from the input space. Furthermore, $\phi$ represents the probability density function and $\Phi$ represents the cumulative distribution function of the standard normal distribution. Following usual notation, $\mu$ and $\sigma$ describe the posterior mean and standard deviation of the \ac{gp} they are conditioned on the observed set $\mathcal{D}$. This $\mu$ used in this context should not be confused with the growth rate from the bioprocess described earlier.

The basic idea of the single-point ($q=1$) \ac{ei} case can be extended into a more general batch setting ($q>1$) to increase throughput, which can be calculated using Monte Carlo sampling \cite{balandat_botorch_2020}.
\begin{equation}
    \text{qEI}_{y^*}(\mathbf{X}|\mathcal{D}) = \frac{1}{N}\sum_{i=1}^{N}\max_{j=1,...,q}\{[\xi^i(\mathbf{x}_j)-y^*]_+\},
\end{equation}
where $\xi^i(\mathbf{x}) \sim f(\mathbf{x})$ are samples from the joint \ac{gp} posterior and $\mathbf{X} = \{\mathbf{x}_1, \dots, \mathbf{x}_q\}$ is the batch of new input points.
It can also be shown that an approximation of the improvement term and a log transformation improves the performance during acquisition function maximization \cite{LogEI_Meta}, yielding
\begin{equation}
    \text{qLogEI}_{y^*}(\mathbf{X}|\mathcal{D}) = \text{log}\int\left(\sum_{j=1}^{q}\text{softplus}_{\tau_0}(f(\mathbf{x}_j)-y^*)^{1/{\tau_\text{max}}}\right)^{\tau_\text{max}}df.
\end{equation}
Here, $\tau_0$ and $\tau_{\text{max}}$ are temperature parameters for the softplus transformation.

The \ac{qlogei} acquisition function is inherently capable of selecting batches $q>1$ but does not provide a mechanism to incorporate data of varying quality and cost. To adjust for that, correction terms can be introduced. These modify the estimated information gain \cite{continuous_fidelity} and make the acquisition function cost-aware \cite{cost_aware_snoek}. This incentivizes the optimizer to prefer evaluations that reflect a well balanced trade-off between information gain and evaluation cost, rather than simply focus on information gain at all costs.
The final acquisition function, including the multi-fidelity correction, takes the following form:
\begin{equation}
    \text{MFqLogEI}_{y^*}(\mathbf{X}|\mathcal{D}) = \frac{\exp{\left[\text{qLogEI}_{y^*}(\mathbf{X}_\bullet|\mathcal{D})\right]}}{c(\mathbf{X})}\cdot \boldsymbol{\rho}(\mathbf{X},\mathbf{X_\bullet}),
\end{equation}
with
\begin{equation}
    \rho_j(\mathbf{x}_j,\mathbf{x}_{j,\bullet}) = \frac{\sigma^2(\mathbf{x}_j,\mathbf{x}_{j,\bullet})}{\sqrt{\sigma^2(\mathbf{x}_j,\mathbf{x}_j)\sigma^2(\mathbf{x}_{j,\bullet},\mathbf{x}_{j,\bullet})}},
\end{equation}
being the correlation between the batch $\mathbf{X}$ and its projection to the target fidelity $\mathbf{X}_\bullet$, which represents the same points in the batch evaluated at the highest (or target) fidelity level. This extension assumes that a batch has only one consistent fidelity level as is further explained in  \Cref{sec:the_bo_workflow}. That way, $\rho$ acts as a penalization term for choosing a lower fidelity level batch over more informative higher level ones (i.e., it corrects for the fact that information from \ac{mtp} experiments for certain points is worth less than experiments from the pilot for the same points). Additionally, $c$ is the evaluation cost depending on the fidelity of the input $\mathbf{X}$. It should be noted that the exponential is applied to bring the acquisition function value of \ac{qlogei} from log-space into regular space before applying cost and fidelity level adjustments.

Normalizing acquisition functions by evaluation cost is a common strategy to promote cost-aware optimization. However, choosing the fidelity level factor is a customized solution, and its impact on the overall optimization has to be carefully evaluated.

\subsubsection{Extended Multi-Point Upper Confidence Bound (Extended qUCB)}
The original \ac{ucb} criterion reflects the fundamental trade-off between exploration and exploitation in \ac{bo} as areas with good mean values but also areas with a lot of uncertainty (and in that way potentially good mean values) are favored through a high acquisition function value. This can be seen in
\begin{equation}
    \text{UCB}(\mathbf{x}|\mathcal{D})=\mu(\mathbf{x})+\beta^{1/2}\sigma(\mathbf{x}),
\end{equation}
where $\beta$ is a hyperparameter that has an effect on the amount of exploration. Similarly to \ac{ei}, \ac{ucb} can be expanded to a general $q>1$ case \cite{qUCB}, which yields an approximate expression computed via Monte Carlo sampling \cite{balandat_botorch_2020}.
\begin{equation}
    \text{qUCB}(\mathbf{X}|\mathcal{D})\approx \frac{1}{N} \sum_{n=1}^{N}\max_{1 \leq j \leq q}(\boldsymbol{\mu}(\mathbf{X}_j)+\sqrt{\beta} \cdot \lvert\mathbf{L}\mathbf{z}_n\rvert_j)
\end{equation} for $\mathbf{z}_n \sim \mathcal{N}(\mathbf{0},\mathbf{1})$
with $\mathbf{L}$ being the Cholesky lower triangular matrix from the decomposition of the posterior covariance matrix $\Sigma$. As with \ac{qlogei}, extensions for cost- and fidelity-awareness were also applied to \ac{qucb}.

\subsubsection{General-purpose Information-Based Bayesian Optimization (GIBBON)}
The \ac{gibbon} acquisition function belongs to the family of information-theoretic, entropy-based acquisition functions \cite{GIBBON}. It provides a computationally efficient lower bound approximation to a general-purpose extension of Max-Entropy Search \cite{wang2018maxvalueentropysearchefficient}. By design, \ac{gibbon} naturally supports batch selection for arbitrary batch sizes and accommodates multiple fidelity levels. Thanks to that, \ac{gibbon} can directly be applied without requiring additional extensions for batching or multi-fidelity scenarios. Following a notation similar to the one introduced by Moss \textit{et al.}~\cite{GIBBON}, the \ac{gibbon} acquisition function GIBBON$(\mathbf{X}|\mathcal{D})$ is defined as: 

\begin{equation} 
\underbrace{\tfrac{1}{2}\log|R|}_{\text{Penalty}} 
- 
\underbrace{
\tfrac{1}{2|\mathcal{M}|} \sum_{m\in\mathcal{M}} \sum_{i=1}^q
\log\left(
1 - \rho_i^2 \cdot \tfrac{\phi(\gamma_i(m))}{\Phi(\gamma_i(m))}
\left[\gamma_i(m) + \tfrac{\phi(\gamma_i(m))}{\Phi(\gamma_i(m))}\right]
\right)
}_{\text{Information gain}}
\end{equation}

In the following, each term is briefly introduced: 
\begin{itemize}
    \item $\mathcal{M} = \{m_1,...,m_n\}$ are samples of the maximum values of the target fidelity objective function $g^*$. In $\gamma(m)$ they are used as a comparison ground over which the improvement is estimated.
    \item $\gamma_i(m)=\frac{m-\mu(\mathbf{x}_{i,\bullet})}{\sqrt{\sigma^2(\mathbf{x}_{i,\bullet},\mathbf{x}_{i,\bullet})}}$ is the standardized improvement with the subscript $\bullet$ indicating that point $\mathbf{x}_i$ is at target fidelity. It is calculated for every combination of maximum value samples in $\mathcal{M}$ and points in the batch of size $q$.
    \item $\phi$ and $\Phi$ are the standard normal probability density and cumulative distribution functions.
    \item $R \in \mathbb{R}^{q \times q}$ is the predictive correlation matrix that captures the interaction between the points in one batch $\mathbf{X}$. Its elements are defined by: $R_{i,j}=\sigma^2(\mathbf{x}_i,\mathbf{x}_j)/\sqrt{\sigma^2(\mathbf{x}_i,\mathbf{x}_i)\sigma^2(\mathbf{x}_j,\mathbf{x}_j)}$. $R$ is calculated for the specific fidelity level of the points in the batch.
    \item $\rho$ is the correlation between the objective value of point $\mathbf{x}_i$ in a batch of a specific fidelity level and the objective value of that point projected to the target fidelity level. It can be computed via $\rho_i = \sigma^2(\mathbf{x}_{i},\mathbf{x}_{i,\bullet})/\sqrt{\sigma^2(\mathbf{x}_{i},\mathbf{x}_{i})\sigma^2(\mathbf{x}_{i,\bullet},\mathbf{x}_{i,\bullet})}$.
\end{itemize}

In the equation for \ac{gibbon}, the first part can be understood as a penalization term based on the predictive correlation matrix of the points in one batch. The second part describes the information gain of each individual point in the batch with respect to every previously sampled potential max value $m \in \mathcal{M}$. Combined, the penalization term corrects an overestimation of the information gain, yielding a lower bound to the information gain. For more details the reader is referred to the original paper \cite{GIBBON}.

\subsection{The Multi-Fidelity Batch Bayesian Optimization Framework}
\label{sec:the_bo_workflow}
This section highlights the implementation of the proposed workflow and used algorithms. All of the explained concepts are implemented in or can be implemented using the Python packages BoTorch~\cite{balandat_botorch_2020} and GPytorch~\cite{gardner_gpytorch_2018}. The framework is capable of handling varying batch sizes, heterogeneous cell clones, and multi-fidelity experimental levels. \Cref{alg:mfbo} outlines the algorithmic structure. As a detail it is important to mention that for the evaluation of $\alpha(\mathbf{X}_\ell)$ in the established fantasies for $\ell \in \{ \text{mid}, \text{high}\}$, different sampling methods are used based on the acquisition function. \ac{gibbon} employs a Gumbel sampling approach introduced in \cite{wang2018maxvalueentropysearchefficient}, while \ac{qucb} and \ac{qlogei} use the already described Monte Carlo sampling approximation.

\begin{algorithm}[H]
\caption{\textbf{Multi-Fidelity Bayesian Optimization with Cost-Aware Acquisition}}
\label{alg:mfbo}
\begin{algorithmic}[1]
\State \textbf{Input:} Number of iterations $N$
\State Initialize dataset $\mathcal{D}_0$ with $n_0$ low-fidelity samples
\For{$n = 1$ to $N$}
    \State Fit multi-task GP on $\mathcal{D}_n$
    \State Sample candidate maximum values from $y^*$
    \For{each fidelity level $\ell \in \{\text{low}, \text{mid}, \text{high}\}$}
        \If{$\ell = \text{low}$}
            \State Find batch $\mathbf{X}_\ell$ maximizing $\alpha(\mathbf{X}_\ell)$ using grid search
        \Else
            \State Find batch $\mathbf{X}_\ell$ maximizing $\alpha(\mathbf{X}_\ell)$ using fantasies
        \EndIf
        \State Compute cost-adjusted acquisition score: $s_\ell \gets \frac{\alpha(\mathbf{X}_\ell) \cdot \rho_{\ell}}{c_\ell}$ 
    \EndFor
    \State Select optimal fidelity and batch $\mathbf{X}^*_{\ell} \in \arg\max_{\ell} s_\ell$
    \State Evaluate selected batch and update dataset: $\mathcal{D}_{n+1} \gets \mathcal{D}_{n} \bigcup \{\mathbf{X}^*_{\ell},y_\ell\}$
    \State Retrain GP model on updated $\mathcal{D}_{n+1}$
\EndFor
\State \Return Best configuration in $\mathcal{D}_N$
\end{algorithmic}
\end{algorithm}

\subsubsection{Individual Treatment of Batches from Different Fidelity Levels}

To reduce the overall complexity of the optimization while still leveraging BoTorch’s routines, the acquisition function optimization was decomposed by fidelity level. That is, for each iteration, a separate batch is optimized for each fidelity level under its respective constraints. These fidelity-specific batches then compete based on their acquisition scores, and the best is selected for evaluation. In this scenario fixed batch sizes for each fidelity level were assumed, as set-ups with standardized sizes are commonly used in laboratories (e.\,g., 48 or 96 wellplates on the \ac{mtp} scale). This comparison simplification can be justified through considering complete batches that were selected based on the same starting conditions in each iteration. This means that the batch improvement measure of each respective acquisition function can be compared even when the batch has different sizes. While the comparison between fidelity levels is no issue, the selection large batches faces computational burdens.

\subsubsection{Microtiter Plates: Dealing with Larger Batch Size and Additional Constraints}

The lowest fidelity level (\acp{mtp}), presents particular challenges due to its larger batch sizes combined with additional constraints. Standard BoTorch methods perform poorly in optimizing this level, since joint optimization of a large cell clone selection, as well as batch-wide process parameters such as temperature and pH (making it a mixed-integer non-linear program), proves computationally infeasible for standard methods. Dedicated large-batch \ac{bo} algorithms have been proposed to address scalability in high-dimensional or large-batch settings \cite{vakili_scalable_2020, nava_diversified_2021}. However, these approaches do not readily accommodate the combination of large-scale categorical diversity (cell clone selection) and problem-specific constraints required here. To overcome this and reduce complexity, a grid search over temperature-pH combinations is conducted. The resulting batches are evaluated for the acquisition function and its values are then used to rank the produced batches for every temperature-pH combination, avoiding repeated, expensive acquisition function evaluations. Furthermore, enforcing the special constraints of the \ac{mtp} scale (see \Cref{sec:optimization_problem}) turns out to be simpler. This is because of the custom implemented grid search method instead of using the limited constraint formulation capabilities of BoTorch.

\begin{algorithm}[ht!]
\caption{\textbf{Grid Search for Low-Fidelity Batch Selection}}
\label{alg:gridsearch}
\begin{algorithmic}[1]
\State \textbf{Input:} Grid of $(T, \mathrm{pH})$ combinations, batch size $q$
\State Initialize best score $s^* \gets -\infty$
\State Initialize best configuration $\mathbf{x}^* \gets \emptyset$
\State Initialize $(F_1, F_2, F_3) \gets (0, F_{max},0)$
\For{each $(T, \mathrm{pH})$ in grid}
    \State Sample $q$ clone types $c_1, \dots, c_q$ using \ac{lhs}
    \State Form batch $\mathbf{X} = \begin{bmatrix} (T, \mathrm{pH}, 0, F_{max}, 0, c_i) \end{bmatrix}_{i=1}^q$
    \State Compute acquisition score $s \gets \alpha^{\text{GIBBON}}(\mathcal{\mathbf{X}})$
    \If{$s > s^*$}
        \State $s^* \gets s$
        \State $\mathbf{X}^* \gets \mathbf{X}$
    \EndIf
\EndFor
\State \Return Best batch $\mathbf{X}^*$
\end{algorithmic}
\end{algorithm}

However, even after this simplification, the clone selection at fixed temperature-pH combinations yielded suboptimal diversity. Especially at low-fidelity levels, where exploration is important, this is unfavorable. The optimizer failed to find acquisition function values that are non-zero for many iterations, which leads to unfavorable point selection. To address this, \ac{lhs} was employed as a \ac{doe} strategy for clone selection, which can be seen in \Cref{alg:gridsearch}. Instead of optimizing to select a diverse batch of cell clones for evaluation, a set of clones is sampled with \ac{lhs} and evaluated combined with the temperature-pH combinations. This also ensures that low-fidelity evaluations remained competitive with the purely \ac{doe}-based industrial baselines. An outline for the grid-search approach is reflected in \Cref{alg:gridsearch}.

\subsubsection{Microbioreactors and Pilot Scale: Leveraging BoTorch Methods}

Higher fidelity levels, due to their smaller batch sizes and relaxed constraints, were amenable to standard BoTorch methods. Internally, BoTorch employs a multi-start L-BFGS-B method for gradient-based optimization \cite{lbfgsb}, combined with a sequential greedy heuristic using fantasy models (also known as Kriging Believer) to construct batches point-by-point (see \cite{kringing_believer} and \Cref{sec:batch_modeling}).

The approach mainly affects \ac{qucb} and \ac{qlogei}, however a similar strategy is applied to the \ac{gibbon} acquisition function. While \ac{gibbon} theoretically can handle the fidelity selection without extensions, for fixed-size batches of purely one fidelity level, it is more practicable to separate acquisition function value optimization and compare batches afterwards. Overall, the key distinction lies in the formulation of the acquisition function itself and the associated preparatory steps, including sampling strategies and pre-computations. In order to end on the target fidelity level and be more comparative to industrial methods, the last evaluation is forced to be on the pilot scale.

\subsection{The Comparison Scenario}
\label{sec:industrial_benchmark}
In order to evaluate the applicability and usefulness of the presented \ac{bo} workflows, they need to be compared with the industrial status quo. Therefore, a benchmark scenario is designed that tries to mimic the funnel-like \ac{doe}-based approach explained in \Cref{sec:intro_challenges-stage}. Due to the reduced number of clones in this study compared to reality (30 instead of 100+), the funnel-based approach needs to be scaled down to be a good benchmark. A well-working mode of operation is found for four iterations on the \ac{mtp} scale, eleven iterations on the \ac{mbr} scale, and five iterations on the pilot-plant scale, leading to a total of 20 iterations.

Starting with exploratory clone sampling at fixed conditions (T, pH) per batch, the best clone from the \ac{mtp} experiments is chosen to go onto the next stage. Here, diverse operating conditions and feeding regimes across the entire search space are tested. From that stage, the best performing point is selected and evaluated within $\pm 5\%$ deviation in the operating conditions to reflect local exploitation on the highest fidelity level. This funnel-like approach is implemented for \acf{lhs}, Factorial Design and Sobol Sequence sampling. Each method selects reaction conditions and cell clones accordingly.

For the random sampling methods (\ac{lhs} and Sobol), candidate points are generated uniformly within the search space of the respective fidelity level. The total number of generated candidates per stage corresponds to the total number of batched experiments planned at that stage. The candidates are then distributed to batches with sizes corresponding to the respective scale and iteratively evaluated. In case of the factorial method, a 3-level full factorial design is generated for each stage. Since not all points can be evaluated due to batch constraints, the generated points are randomly shuffled and grouped into batches according to the batch size for that stage, until the total number of batches required is reached.

\section{Results \& Discussion}
\label{sec:results_and_discussion}
\subsection{The Optimization Problem}
\label{sec:optimization_problem}
\begin{figure*}
    \centering
    \includegraphics[width=1\linewidth]{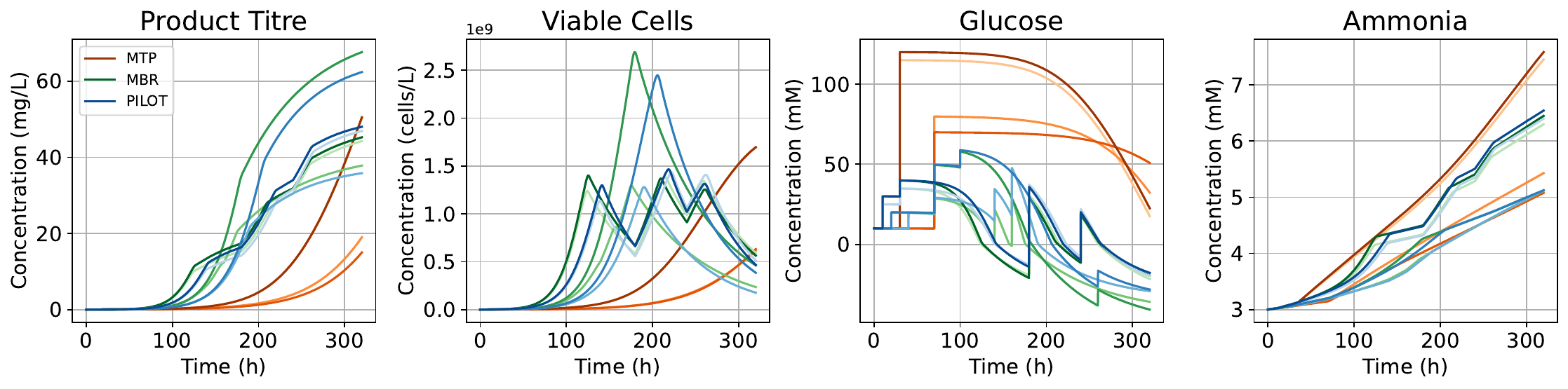}
    \caption{\textbf{Runs for simulated experiments on \ac{mtp}, \ac{mbr} and pilot scale for one clone}\\
    \ac{mtp} (orange), \ac{mbr} (green) and pilot (blue) are compared for the same clone, while the feeding regime and the temperature-pH combinations were varied, which is indicated by different shades of the respective reactor scale colour. This setup yields a variety of concentration profiles for the product titer, the viable cell density (a measure for the amount of cells that are alive in the reactor), as well as the substrate concentrations of glucose and ammonia. Especially the high variety of end point concentration values for the product titer showcases the suitability for a hard to optimize objective function.}
    \label{fig:process_dynamics}
\end{figure*}

This work examines case studies based on \ac{cho} cells producing an arbitrary product molecule, inspired by common industrial applications such as antibody production~\cite{CHO_review, CHO_review_2}. Here, challenges include highly non-linear dynamics, measurement noise, and scale-dependent behavior across multiple experimental fidelity levels. The model introduced in \Cref{sec:bioprocess_model} reflects these features while remaining sufficiently simple to enable a systematic feasibility study. Conclusively, the result is a model that showcases complex process dynamics (see \Cref{fig:process_dynamics}). The goal was not to construct a detailed process model for a specific bioprocess, but rather to develop a computationally tractable framework that retains the essential complexity and structural characteristics of realistic systems.

To reduce model complexity and focus on the most practically relevant performance metric, the accumulated product titer at the end of a fixed experiment duration is used as the scalar objective in the optimization. The consideration of alternative or additional time-resolved metrics, such as intermediate titers or peak production rates, is left for future work. This system with all of its characteristics can now be used as an objective function in the formal optimization problem.

The goal is to maximize a black-box objective function $ g(\mathbf{x},c)$, where $\mathbf{x} \in \mathbb{R}^5$ denotes the experimental configuration:
\begin{equation*}
   \mathbf{x} = [T, \mathrm{pH}, F_1, F_2, F_3]^\top
\end{equation*}
\begin{table}[h]
  \centering
  \caption{\textbf{Variables and their descriptions used in the model}}
  \label{tab:variables_description}
  \begin{tabular*}{0.48\textwidth}{@{\extracolsep{\fill}}cc}
    \hline
    \textbf{Variable} & \textbf{Description} \\
    \hline
    $T$ & Temperature \\
    $\mathrm{pH}$ & Acidity level \\
    $F_1, F_2, F_3$ & Substrate feeding amounts \\
    $c$ & Clone type (categorical) \\
    \hline
  \end{tabular*}
\end{table}

subject to the following constraints:
\[
\begin{aligned}
\max_{\mathbf{x},c} \quad & g(\mathbf{x},c) \\
\text{s.t.} \quad 
& \sum_{i=1}^{3} F_i \leq F_{\text{max}} \\
& 30 \leq T \leq 40, \quad 6 \leq \mathrm{pH} \leq 8 \\
& 0 \leq F_i \leq F_{\text{max}}, \quad \text{for } i = 1, 2, 3 \\
& c \in \mathcal{C}
\end{aligned}
\]

Here, $ F_{\text{max}}$ denotes the total allowable substrate amount added across all feed pulses, and $\mathcal{C}$ is the discrete set of available cell clones.

The bioprocess can be evaluated at three different reactor scales, each corresponding to a fidelity level in a multi-fidelity optimization framework. The pilot scale represents the highest fidelity and serves as the objective function, denoted by $g_{\text{high}}(\mathbf{x},c)~=~~g(\mathbf{x},c)$. The two lower-fidelity evaluations correspond to the \ac{mbr} and \ac{mtp} scales, denoted by $g_{\text{mid}}(\mathbf{x},c)$ and $g_{\text{low}}(\mathbf{x},c)$, respectively.

Due to physical and procedural limitations, experiments at the smallest scale (\ac{mtp}) are subject to additional constraints. In practice, it is not possible to vary the temperature between wells within the same \ac{mtp} batch. Similarly, pH values are typically fixed per plate due to technical limitations. Moreover, the complexity of preparing \ac{mtp} experiments, which is often done manually and with limited automation, makes it impractical to test diverse feeding regimes. To account for these restrictions, no feeding is assumed for \ac{mtp} experiments in the main study ($F_\text{max}=0$), however variations on this can be found in \Cref{sec:flexible_feeding}. As a result, the \ac{mtp}-level decision space is limited to fixed combinations of $T$, $\mathrm{pH}$, and a selected subset of clones from $\mathcal{C}$. In contrast, the two higher scales (\ac{mbr} and pilot) are not subject to these constraints and allow for full exploration of the search space defined above. 

To address this optimization problem, we applied a multi-fidelity batch \ac{bo} workflow that explicitly incorporates the fidelity structure of bioprocess experimentation and allows simultaneous evaluation of multiple experimental conditions. By integrating information across fidelity-levels, this approach improves optimization efficiency and supports more informed decision making under experimental constraints (see \Cref{sec:the_bo_workflow}).

The workflow’s performance was assessed in several case studies designed to reflect industrially relevant scenarios. After outlining these case studies and highlighting their key variations, we present the most pertinent results. Across all studies, two main factors were varied: the underlying distribution of clone performance across scales and the representation of clone attributes in \ac{bo}.

\subsection{Results of the Clone Representation Case Study}
In industrial practice, process models developed alongside wet-lab experimentation rarely include detailed representations of clone characteristics. In this study, we evaluated two alternatives to the commonly used one-hot encoding: \ac{icm} \cite{multi_task_1} and entity embedding \cite{entity_embedding}, both described in \Cref{sec:categorical_variables}. The aim was to determine which approach is most effective for the transfer learning task. The \ac{icm} method uses a kernel that directly learns covariances between clones. In contrast, the entity embedding method maps clones into a Euclidean space, where distances between points reflect similarities in clone behavior. From this embedding, covariance-like quantities can be derived. In our application, these relationships were inferred solely from experimental data, without prior knowledge, although additional information such as genetic data could be incorporated to improve the model. More complex information structures have already been used as features in optimization problems \cite{gomez-bombarelli_automatic_2018, moss_boss_2020}.

\Cref{tab:clone_representation} presents the results for varying the clone representation across the acquisition functions introduced in \Cref{sec:acquisition_functions}. The comparison reveals that certain acquisition functions perform better overall and exhibit greater sensitivity to the choice of clone representation. For \ac{qucb}, performance shows notable changes depending on the representation, with the two-dimensional entity embedding achieving the best results (\SI{61}{\milli\gram\per\liter}). There are also differences for \ac{qlogei}, where the \ac{icm} kernel turns out to be about \SI{14}{\percent} better than the two-dimensional entity embedding, and about \SI{7}{\percent} better than the three-dimensional one. Lastly, using a three-dimensional entity embedding with the \ac{gibbon} acquisition function increases the average termination value by about \SI{2}{\percent} relative to the two-dimensional embedding and by \SI{22}{\percent} relative to the \ac{icm} kernel.

\begin{table}[h]
  \centering
  \caption{\textbf{Comparison of different clone representation strategies}\\
  The table presents the final cumulative maximum endpoint concentrations (mean and standard deviation) for three acquisition functions and varying clone representations across 10 runs for one of the two tested clone distributions (Distribution A). The highest performing representation for each acquisition function is highlighted in bold.}
  \label{tab:clone_representation}
  \begin{tabular*}{0.7\textwidth}{@{\extracolsep{\fill}}llcc}
    \hline
    Method & Clone Representation & Mean (mg/L) & Std (mg/L) \\
    \hline
    \ac{gibbon} & \ac{icm} & 54 & $\pm$ 15 \\
     & Entity Embedding (2D) & 65 & $\pm$ 14 \\
     & \textbf{Entity Embedding (3D)} & \textbf{66} & \textbf{$\pm$ 12} \\
    \ac{qlogei} & \textbf{\ac{icm}} & \textbf{58} & \textbf{$\pm$ 12} \\
     & Entity Embedding (2D) & 51 & $\pm$ 16 \\
     & Entity Embedding (3D) & 54 & $\pm$ 11 \\
    \ac{qucb} & \ac{icm} & 53 & $\pm$ 11 \\
     & \textbf{Entity Embedding (2D)} & \textbf{61} & \textbf{$\pm$ 12} \\
     & Entity Embedding (3D) & 56 & $\pm$ 18 \\
    \hline
\end{tabular*}
\end{table}

While not always being the ideal choice, these results suggest that the three-dimensional entity embedding captures meaningful structure that all \ac{bo} version can exploit to distinguish well-performing clones from unsuitable ones. One might expect that generally increasing the embedding dimensionality could further enhance algorithmic performance; however, the tests conducted indicate otherwise (see \Cref{sec:higher_dimensions}). The key trade-off in using embeddings in this setting lies between the level of detail encoded in the latent space and the feasibility of learning that space with limited data. Especially for \ac{qucb} and \ac{gibbon}, a latent representation appears advantageous while remaining learnable within the data constraints. This aligns with the findings of Hutter et al. \cite{entity_embedding}, who reported that a low-dimensional embedding space was sufficient to minimize modeling errors when representing cell characteristics with a \ac{gp}, with no significant additional benefit from higher dimensionality.

Overall, the combination of \ac{gibbon} with a three-dimensional entity embedding achieved the best performance among the \ac{bo} approaches tested. 

\subsection{Results of the Cell Distribution Case Study}
\begin{figure}[ht!]
    \centering
    \includegraphics[width=1\linewidth]{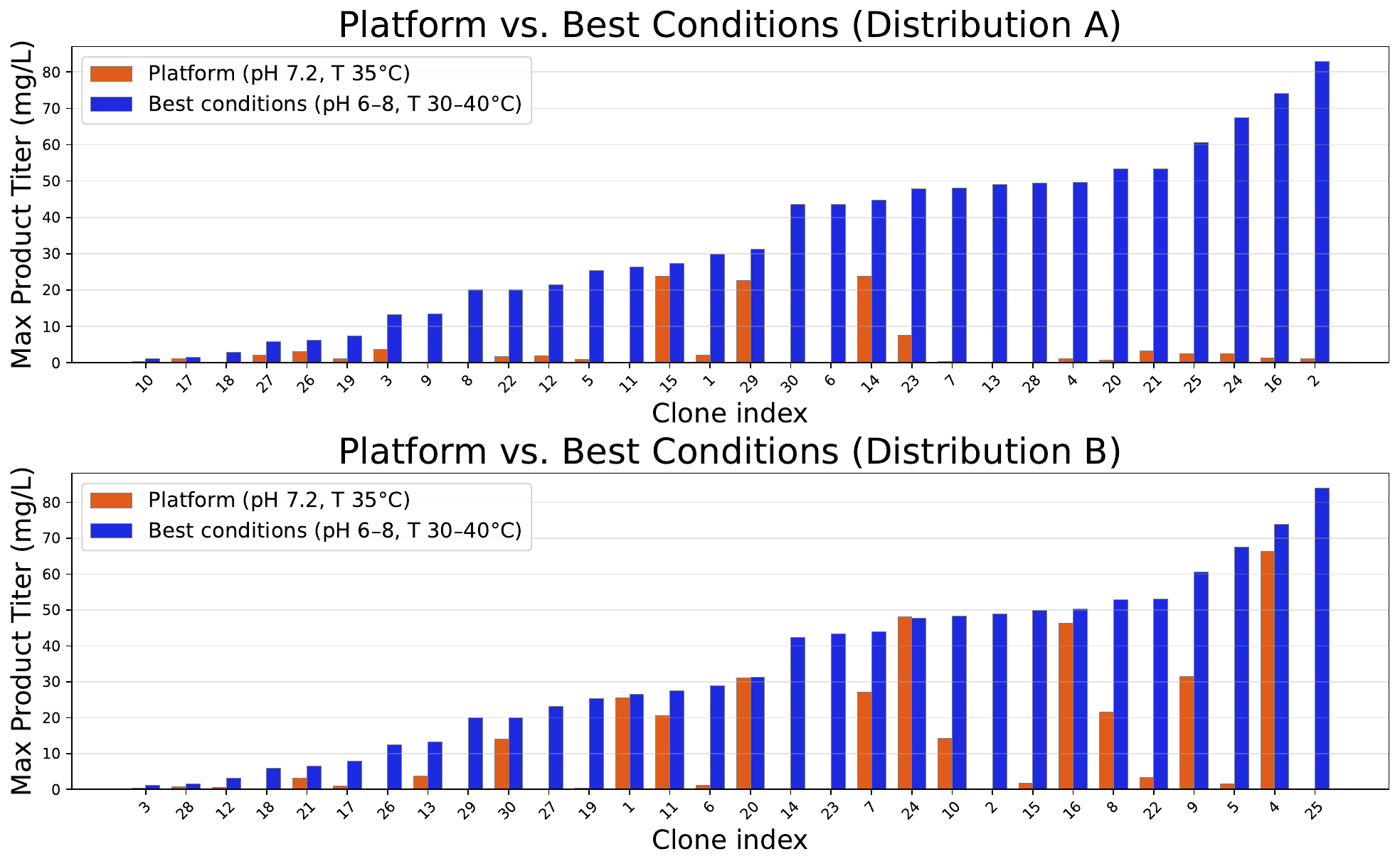}
    \caption{\textbf{Ordered maximum results for every clone and fidelity}\\
    The figure shows two distributions of clones and the final product titer for every clone on the pilot scale under platform conditions (red) and the individual ideal conditions (blue). While for Distribution~A (upper plot) clear differences in best performance under different conditions can be seen, the performance distribution is much more aligned between platform conditions and ideal conditions for Distribution~B (lower plot).}
    \label{fig:clone_distribution}
\end{figure}

In addition to learning relationships between clones, the underlying distribution of clone performance plays a key role, as it reflects differing goals and prior knowledge during process development. Depending on the development context, these distributions can vary considerably, thereby influencing the search strategy. Two such distributions for 30 cell clones are shown in \Cref{fig:clone_distribution}, each with distinct characteristics. Distribution~A represents a scenario with unmapped scale-up behavior, where clone performance under platform conditions diverges substantially from performance under each clone's individual optimized process conditions. Such a distribution is likely when entirely novel processes are developed or when the platform conditions used during screening are poorly matched to the conditions later found to be optimal, meaning the unrepresentative screening data often misleads traditional selection methods. In contrast, Distribution~B represents a more mature process development stage, where clone behavior under platform conditions correlates stronger with outcomes encountered under optimized process conditions. Since MTP-scale screening is conducted under platform conditions, the lower fidelity experiments yield informative priors for the optimization process under this distribution, making it much simpler to select promising candidates that perform well at the larger scale.

In this latter case of Distribution~B, a more balanced trade-off between exploration and exploitation may be advantageous. While these informative priors make it straightforward to identify promising candidates, relying too heavily on immediate exploitation risks an early lock-in into a merely good solution rather than the true global optimum. Based on these two representative cases, which plausibly cover the spectrum of real process development, we conducted a benchmark of the proposed framework against the industrial scenario described in \Cref{sec:industrial_benchmark}.

\begin{figure*}[ht!]
    \centering
    \includegraphics[width=1\linewidth]{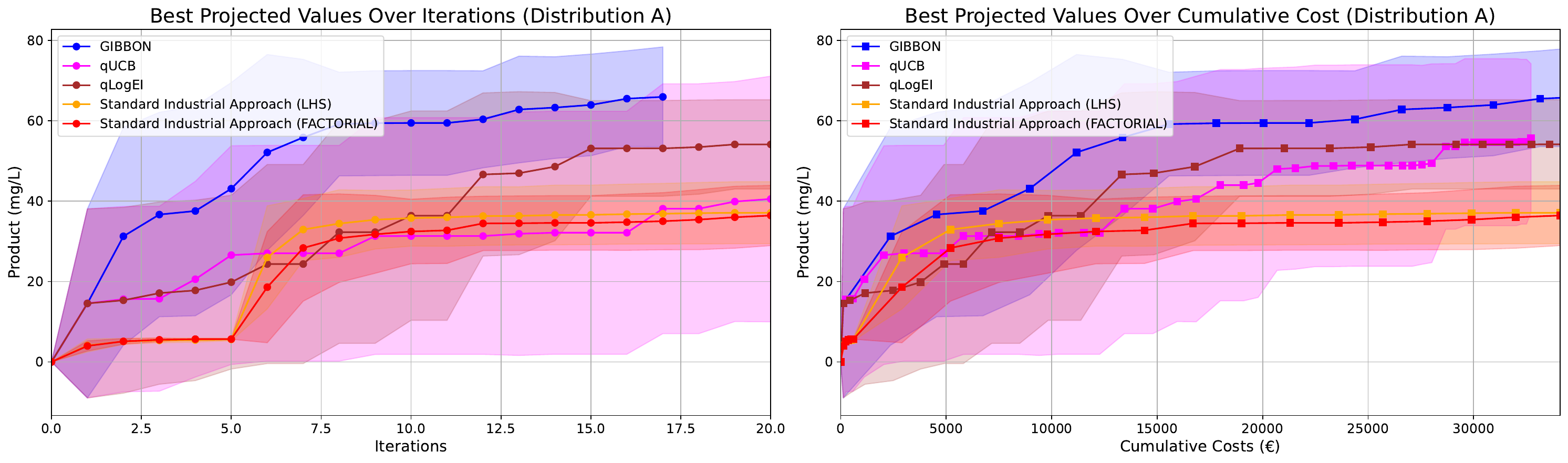}
    \caption{\textbf{Comparison of BO results for Distribution~A} \\
    Showing the best found projected values for multiple optimization approaches. This plot compares the performance of \ac{gibbon} (blue), \ac{qlogei} (brown), \ac{qucb} (magenta), the industrial approaches with factorial design (red) and Latin Hypercube sampling (yellow). The shaded areas represent the empirical standard deviation from the mean values depicted by the lines. For this scenario a distribution of clones with more misaligned scale-behavior was used (see Distribution~A in \Cref{fig:clone_distribution}). On the left side the behavior is plotted over the first 20 iterations while on the right it is shown over cumulative cost of evaluation up to €34,100. The plots enable a comparison of the optimization behavior within the iteration and cost boundaries of the industrial benchmark.}
    \label{fig:results_dist_A}
\end{figure*}

\Cref{fig:results_dist_A} presents the results for Distribution~A, where the \ac{bo} workflow, which is tested with three different acquisition functions, is compared with the industrial funnel-like approach using multiple sampling strategies. The comparison was conducted over a limit of 50 iterations and €34,100 cumulative cost for the \ac{bo} variants, and the mean and standard deviation across 10 independent runs are reported. For the \ac{bo} approach, the three-dimensional entity embedding was selected, as it yielded the best results for this distribution.

For Distribution~A, the optimization task primarily revolves around when to use \ac{mtp}-scale experiments for exploitation. In this challenging scenario, \ac{bo} acquisition functions were able to successfully navigate the search space to locate the well-performing regions, setting them ahead of \ac{lhs} sampling and factorial design. A notable feature in the evaluation profile is the distinct performance jump observed for the industrial methods after the 5th iteration. This step-change occurs as the feeding data from the \ac{mbr}-scale experiments are integrated, projecting the selected candidates onto the pilot-scale tracking space to map the evolution of the objective function. Despite this baseline lift, \ac{gibbon} achieves the highest average objective function value of \SI{66}{\milli\gram\per\liter} at a cost of €35,260 (\SI{3}{\percent} over benchmark) in 17 iterations. This is closely followed by \ac{qlogei} with \SI{54}{\milli\gram\per\liter} at a cost of €35,304 (\SI{4}{\percent} over the benchmark cost) in 27 iterations. Notably, both \ac{gibbon} and \ac{qlogei} manage to surpass the maximum value of the best industrial baseline method remarkably early, doing so at iterations 4 and 12, respectively. This represents a significant objective value improvement, being up to \SI{78}{\percent} higher than the best industrial \ac{doe} method (\ac{lhs} at \SI{37}{\milli\gram\per\liter}). \ac{qucb} eventually terminates with an objective value of \SI{56}{\milli\gram\per\liter} at a total cost roughly \SI{4}{\percent} lower than the industrial baseline. Surprisingly, the \ac{qucb} acquisition function struggles in the early stages compared to the industrial setup, showing only little improvement at the standard industrial length of 20 iterations. Nevertheless, it surpasses the benchmark at iteration 17 and continues to improve over the maximum allowed 50 iterations. In general, it is apparent the evaluated methods tend to converge toward a stable performance level after their initial evaluations, with only smaller improvements observed thereafter.

\begin{figure}[ht!]
    \centering
    \includegraphics[width=1\linewidth]{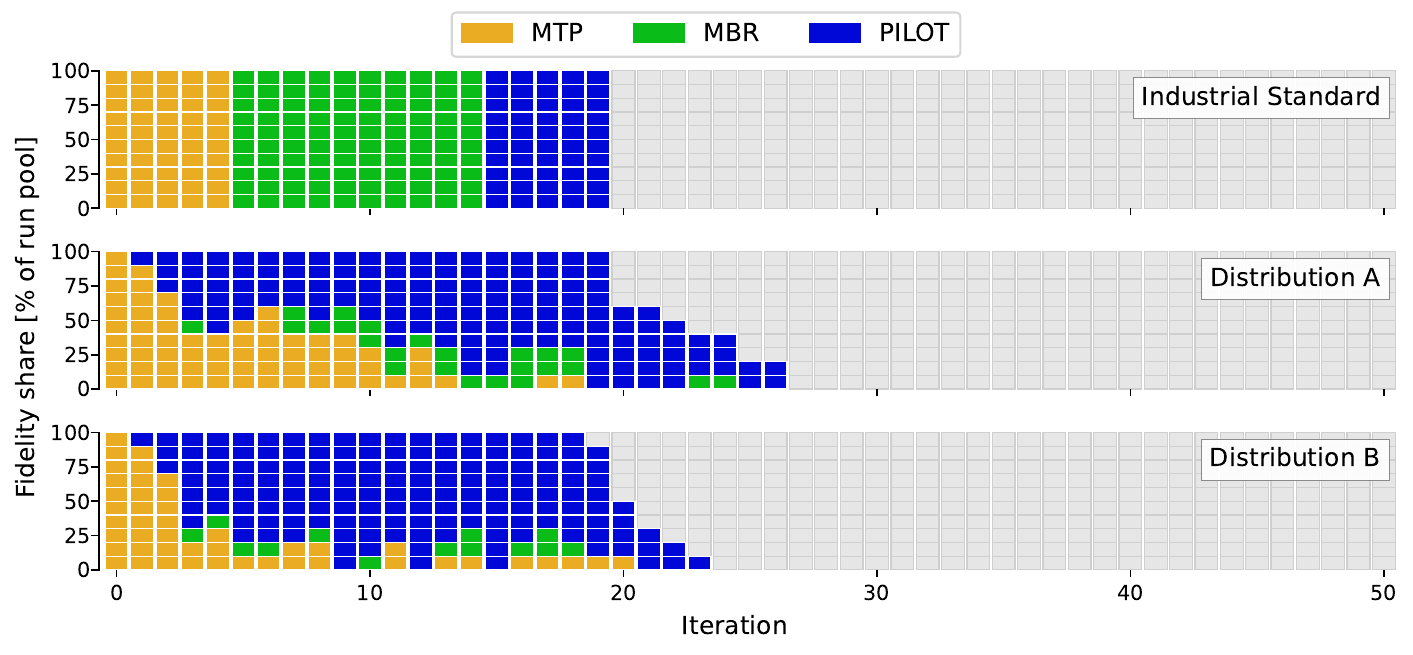}
    \caption{\textbf{Overview of used fidelity levels throughout BO iterations with \ac{qlogei}} \\
    The plot illustrates the fraction of evaluations conducted at each fidelity level per iteration for Distribution~A and B in the two bottom rows. The top row shows the industrial setup for comparison. Initial iteration consist exclusively of \ac{mtp} evaluations (yellow), which can be followed by a mixture of scales with \ac{mbr} (green) and pilot experiments (blue). Between these stages, a gradual shift is observed from \ac{mtp} to \ac{mbr} (blue) and mixing in \ac{mbr} evaluations after 3 iterations. No run reached the maximum iteration count of 50, with the longest run taking 25 iterations.}
    \label{fig:fidelity_distribution_qlogei}
\end{figure}

In order to further analyze the decision making of the acquisition functions, the selected fidelity levels per iteration can be visualized. This gives an understanding on which fidelity levels the algorithm puts a focus on and at which stage. When examining the fidelity evaluation distribution of \ac{qlogei} (see \Cref{fig:fidelity_distribution_qlogei}), it exhibits a systematic exploration pattern that strikingly resembles the methodical approach of industrial practices. Initially starting with \ac{mtp} observations and switching to higher fidelity evaluations in later iterations. This result underscores the validity and effectiveness of strategies often developed through extensive practical experience in industry. However, it also provides a strong indication that when employing transfer learning methods like multi-fidelity \ac{bo}, it is beneficial to evaluate higher-fidelity levels earlier in the process. Such early evaluations provide crucial information regarding the interdependencies between different reactor scales, leading to an improved selection of experiments at all levels. Generally speaking, this means that low-fidelity information combined with strategically timed high-fidelity evaluations (not just at the latest stage) can be highly beneficial. While this demonstrates an interesting proof of concept, an important consideration for future applications is the incorporation of additional constraints related to quality and approvals, such as the mandatory clinical stages encountered in the pharmaceutical industry. Several approaches could be envisioned for integrating the necessary trials into the optimization process. One possibility is the use of a custom dynamic cost function, which varies the cost of fidelity levels to softly guide the algorithm toward the scales that are relevant for the stage in question. Another strategy is to treat the results and inputs of these specialized experiments as regular data points, thereby enriching the dataset and ensuring that information from every experiment contributes to the surrogate model. If the experiments take place at a different unseen scale, they may either be aligned with the fidelity level that is closest to their characteristics or, alternatively, motivate the introduction of a new fidelity level that is qualitatively positioned between existing levels on the scale. Additional details regarding the distribution of fidelity level evaluations across the acquisition functions are provided in \Cref{sec:fid_level_dist}.

\begin{figure*}
    \centering
    \includegraphics[width=1\linewidth]{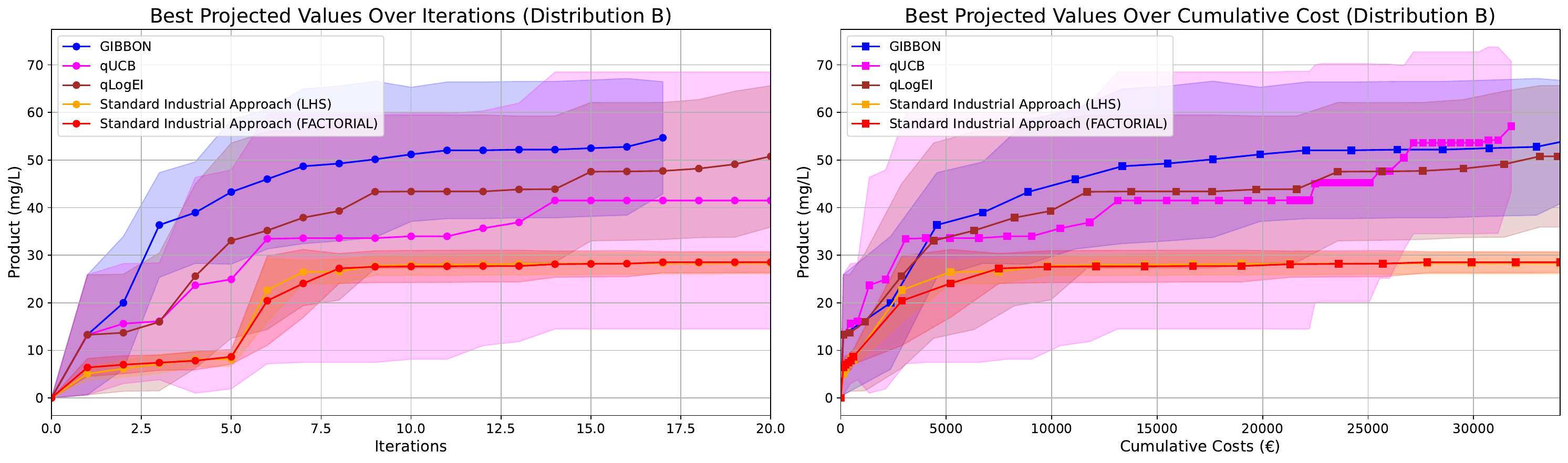}
    \caption{\textbf{Comparison of BO results for Distribution~B} \\
    Showing the best found projected values for multiple optimization approaches. This plot compares the performance of \ac{gibbon} (blue), \ac{qlogei} (brown), \ac{qucb} (magenta), with the industrial approaches \ac{lhs} (yellow) and factorial design (red). For this scenario a distribution of clones with higher behavior across scales was used (see Distribution~B in \Cref{fig:clone_distribution}). On the left side the behavior is plotted over the first 20 iterations while on the right it is shown over cumulative cost of evaluation up to €34,100. The plots enable a comparison of the optimization behavior within the iteration and cost boundaries of the industrial benchmark.}
    \label{fig:results_dist_B}
\end{figure*}

For Distribution~B, the overall trends remain qualitatively similar (see \Cref{fig:results_dist_B}), with all \ac{bo} variants demonstrating significantly stronger performance compared to conventional approaches across all iterations. However, showing a strong relative improvement, \ac{qucb} achieves the highest overall peak value of \SI{57}{\milli\gram\per\liter} over 50 iterations, while \ac{gibbon} reaches \SI{55}{\milli\gram\per\liter} over 17 iterations and \ac{qlogei} reaches \SI{51}{\milli\gram\per\liter} over 24 iterations. Crucially, all three \ac{bo} methods easily surpass the final product concentration of the best \ac{doe} method, factorial design, which yielded a peak of \SI{29}{\milli\gram\per\liter}. As observed in the previous scenario, a similar performance jump for the industrial methods occurs after the 5th iteration due to the integration of the \ac{mbr} feeding data projection. Nevertheless, the \ac{bo} variants consistently maintain their lead and exceed the final industrial benchmark peak remarkably early in the workflow: \ac{gibbon} after only 3 iterations, \ac{qlogei} after 5 iterations, and \ac{qucb} after 6 iterations. These even earlier gains, compared to Distribution~A, highlights the effectiveness of the framework to exploit the informative priors provided by the better-aligned screening scale in Distribution~B. This is further underlined by the strong improvement of the \ac{qucb} behavior.
Given that the standard industrial benchmark length was limited to 20 iterations, these results demonstrate the potential of \ac{bo} to substantially compress development timelines while maximizing yields. Looking at this 20-iteration mark, \ac{qlogei} achieves a \SI{78}{\percent} objective value increase at a \SI{3}{\percent} cost reduction, while \ac{qucb} delivers a \SI{100}{\percent} performance increase alongside a substantial \SI{40}{\percent} cost decrease. Meanwhile, \ac{gibbon} secures its peak terminal value right at the 17-iteration mark, representing a \SI{92}{\percent} improvement over the factorial design baseline, albeit at a higher cost.

It is important to note that the \ac{doe} methods operate with fixed costs of €34,100, as they are designed to perform a specific amount of iterations on each fidelity level. Furthermore, when looking at the uncertainty of methods, one can clearly see the larger variance in the plots for \ac{bo} methods. This could be explained by higher exploration, which as an effect leads to higher variance in results.

Looking conclusively at the implementation of cascaded screening strategies in industrial practice, a clear trade-off with distinct advantages and structural limitations can be observed. In the standard, unsupplemented \ac{mtp} screening phase, the lack of a feeding regime leaves the \ac{mtp} scale to rely entirely on the initial substrate concentration. Generally speaking, conventional industrial methods fundamentally fail to find good clones without this feeding information at the \ac{mtp} scale. For performance profiles characteristic of Distribution~A, this traditional funnel-like approach induces a critical early lock-in effect. Interestingly, when transitioning from Distribution~A to the better-aligned Distribution~B, industrial methods exhibit a strong empirical standard deviation reduction of approximately \SI{30}{\percent} on average across runs and iterations. However, this drop in variance does not equate to improved performance. Regardless of how well-aligned the \ac{mtp} information is to the pilot scale, the industrial methods simply fail to identify high-performing clones in both scenarios. Because the unsupplemented \ac{mtp} phase is highly uninformative in the beginning, the lock-in effect dictates the first screening stage. Since there is no clone exploration in the later stages, these subsequent \ac{mbr} and pilot-scale phases are strictly limited to the exploitation of the prematurely selected, suboptimal clones, failing to carry forward any valuable early condition information.

In contrast, the application of \ac{bo} successfully overcomes these sequential bottlenecks. By mixing the fidelity levels dynamically instead of relying on a strictly staged filter, and by continuously incorporating the information of all previous experiments into the \ac{gp} surrogate model, the \ac{bo} framework manages to actively explore the clone library and the process conditions simultaneously. This holistic knowledge usage allows \ac{bo} solutions to perform robustly regardless of the underlying distribution. Within this framework, the \ac{qucb} acquisition function profits significantly from Distribution~B, where the scales are better aligned. This efficiency makes intuitive sense, as \ac{qucb} possesses an inherent tendency to select low-scale experiments and under aligned conditions, these inexpensive \ac{mtp} runs become more informative regarding late-stage pilot behavior. Consequently, the multi-fidelity approach deployed in the earlier iterations compensates for cross-scale discrepancies, clearly underline by higher average endpoint concentrations and a more robust exploration trajectory.

This dynamic shifts, however, in a non-standard scenario featuring a flexible feeding strategy at the screening scale (e.g., in an automated lab environment capable of delivering targeted feeding pulses per \ac{mtp} batch), as discussed in \Cref{sec:flexible_feeding}. In contrast to relying solely on the initial solution concentration, flexible feeding in the \ac{mtp} scale represents the opposite operational extreme. This setup provides data regarding the kinetic feeding responses of the clones, thereby offering indicators of overall clone quality and growth behavior. Yet, while this extra information establishes valuable priors for optimization, it can also introduce more complex, potentially misleading signals if the scale-down models are sub-optimally calibrated as also can be seen in \Cref{sec:flexible_feeding}.

While traditional industrial methods risk getting stuck at a supposedly good clone too early under this complex configuration, certain acquisition functions, such as \ac{gibbon}, face the opposite problem. Analyzing the optimization behavior of \ac{gibbon} reveals a strong, systematic tendency toward high-fidelity experiments (see \Cref{fig:fidelity_distribution_gibbon}). It can be hypothesized that the constrained \ac{mtp}-scale selection format does not naturally align with how \ac{gibbon} would select batches. The batch selection process is more like selection process rather than natively selected through continuous optimization. The information-theoretic diversity term within the \ac{gibbon} acquisition function (see \ac{gibbon} in \Cref{sec:acquisition_functions} heavily penalizes these \ac{mtp} batches. Consequently, \ac{gibbon} might score them poorly relative to the higher-fidelity level batches, which are selected natively by optimizing the acquisition function. This exploration-exploitation imbalance may over-prioritize expensive experiments. Therefore, some care must be taken with the fidelity selection mechanisms when deploying \ac{gibbon}.

Ultimately, despite these specific sensitivities under non-standard screening, the mixed-fidelity \ac{bo} approach proves robust against the architectural limitations of conventional industrial screening pipelines and can be recommended for real-world process development.

\subsection{Overarching Recommendations and Future Directions}
Based on these findings, specific recommendations can be provided regarding which \ac{bo} methods perform well under different conditions. This can help choosing the right configuration and potentially fine-tune the workflow based on prior knowledge of clone distributions or scale correlations.

If the primary goal is maximizing final yield, \ac{gibbon}, when combined with the three-dimensional entity embedding, demonstrates the strongest capacity to outperform other methods, though this superior objective value comes at a higher cumulative cost and a strong tendency to select high-fidelity experiments. For scenarios requiring a more balanced operational profile, \ac{qlogei} emerges as a compelling choice; by encouraging a more intensive mixing of fidelity levels, it effectively reduces overall scale-up costs without sacrificing optimization efficacy. Meanwhile, if cost reduction is the guiding constraint, \ac{qucb} proves highly reliable, successfully surpassing traditional benchmarks even in uninformative settings like Distribution~A while minimizing the use of expensive higher-fidelity evaluations.
Ultimately, the integration of these \ac{bo} methods into experimental workflows offers an improvement over the industrial status quo, allowing practitioners to select an acquisition function tailored specifically to their budget and performance priorities.

To the best of our knowledge, this is the first study to systematically compare multi-fidelity batch \ac{bo} with the industrial approach to bioprocess development for \ac{cho} cells. Using a simulation study, we demonstrated that parallel process scale-up and optimization are feasible and can improve upon the current industrial practice. Moreover, the results show that the cost of experimentation can be reduced substantially, further highlighting the importance and potential of multi-fidelity batch \ac{bo} in biotechnology. 

In the future, several small adjustments to the bioprocess simulation could be made to increase its realism. One consideration is that the bioprocess model used here may be more challenging to optimize than its real-world counterpart, as the \ac{mtp} scale was modeled to be relatively uninformative. This modeling choice likely results in lower correlations than those typically observed between real \acp{mtp} and pilot scales (see, for example, the effective \ac{mtp} data generation and its utilization in \cite{HTP_MTP_platform}). Adjusting the scale parameters could make the problem more realistic by ensuring more similar behavior across scales. In reality, this would reflect well-initialized experiments (e.g., with tighter controls or more precise measurements) which can improve the transfer learning capability of multi-fidelity and multi-task \acp{gp}, supporting better optimization.

Furthermore, it remains a pertinent question whether the current modeling of fidelity levels is sufficient for the underlying \ac{gp} to accurately capture the differing behaviors across scales. An alternative modeling approach, such as employing a multi-task kernel (similar to the one used for clone modeling), could be explored. However, depending on the specific method, this might introduce additional hyperparameters, making the learning process more challenging. In contrast, the method presented here remains relatively simple, relying on only a single learned parameter, which is the lengthscale for the fidelity within the automatic relevance detection \ac{rbf} kernel. Beyond understanding the correlations between discrete fidelity levels, a significant opportunity lies in the dynamic learning of the fidelity levels themselves and their associated quality. Instead of using a predefined set of values, the approach would allow for a continuous spectrum of fidelity levels. This could enable a more accurate evaluation of experimental outcomes and the information content they provide, ultimately leading to a more robust and adaptable optimization framework.

To further enhance the efficacy and precision of batch selection, particularly given the mixed-variable and highly non-linear characteristics of the acquisition function landscape, the development of a more sophisticated optimization algorithm for this set-up could be considered. Such an algorithm could especially improve the diversity of large batches on the lowest fidelity.

Beyond improved decision making, integrating more flexible \acp{gp}, such as those using the Linear Model of Coregionalization kernel~\cite{LMC_kernel}, may further improve the learning of complex clone behavior. This could lead to a better model fit and more informed decision making through the acquisition function. With better behavioral modeling, the number of clones considered in the workflow could be scaled up. However, this would necessitate the use of sparse \ac{gp} techniques to manage the computational load associated with larger datasets, especially in real-world scenarios involving hundreds of clones.

\section{Conclusion \& Outlook}
Bioprocesses hold immense potential for the sustainable production of pharmaceuticals and specialty chemicals. Accelerating bioprocess development while minimizing experimental costs is essential to enable scalable and efficient manufacturing of new biological products. This work investigated multi-fidelity batch \acf{bo} workflows as a means to streamline bioprocess development through simulation-based evaluations.

In this work a \ac{bo} framework is introduced, incorporating batch selection and fidelity modeling, and evaluated using a custom-designed bioprocess simulator. This simulator captured key challenges of real-world processes, including non-linear dynamics, scale-dependent variability, and heterogeneous clone performance distributions. It thus provided a very challenging environment for systematically comparing optimization strategies.

Our results revealed that, compared to standard industrial strategies, the proposed \ac{bo} framework achieved superior performance in scenarios with differences in clone performance over scales, as well as scenarios with more aligned clone performance, provided that the clone representation model was carefully chosen (see \Cref{sec:results_and_discussion}). In particular, the \ac{gibbon} and \ac{qlogei} acquisition functions performed strongly under these conditions and we showed that a simultaneous cost reduction and optimization performance increase could be achieved.

This work contributes to the growing body of research exploring machine learning-driven approaches for bioprocess optimization. While potentials remain, especially in terms of scaling and real-world applicability, the outlook for automated, data-efficient bioprocess development is promising. Such methods are becoming increasingly important for rapidly designing high-performing and sustainable bioprocesses that advance biopharmaceutical innovation, specialized molecule production, and affordable value chemicals.

While the current framework targets the optimization of single processes, future opportunities may lie in the reuse of individual components. In particular, transfer of surrogate models and optimization trajectories from one bioprocess to another could accelerate development and reduce experimental burden. Capturing fundamental characteristics across scale-ups and optimization runs would be central to enabling such cross-process transfer learning. To assess the feasibility of this idea and further validate the framework, applying the full pipeline to real-world bioprocess development data represents a crucial next step.

\section*{Abbreviations}
\begin{acronym}[GIBBON]
\setlength{\itemsep}{-\parsep}
\acro{bo}[BO]{Bayesian optimization}
\acro{cho}[CHO]{Chinese hamster ovary}
\acro{ddo}[DDO]{data-driven optimization}\acro{gp}[GP]{Gaussian process}
\acro{doe}[DoE] {Design of Experiments}
\acro{ei}[EI]{expected improvement}
\acro{gibbon}[GIBBON]{General-purpose Information-Based Bayesian OptimizatioN}
\acro{icm}[ICM]{intrinsic coregionalization model}
\acro{lhs}[LHS]{Latin hypercube sampling}
\acro{mbr}[MBR]{microbioreactor}
\acro{mtp}[MTP]{microtiter plate}
\acro{ode}[ODE]{ordinary differential equation}
\acro{qlogei}[qLogEI]{multi-point log transformed expected improvement}
\acro{qucb}[qUCB]{multi-point upper confidence bound}
\acro{rbf}[RBF]{radial basis function}
\acro{ucb}[UCB]{upper confidence bound}

\acrodefplural{gp}[GPs]{Gaussian processes}
\end{acronym}

\section*{Author contributions}
\textbf{AM:} Investigation; Formal Analysis; Methodology; Software; Visualization; Writing – Original Draft; Writing – Review \& Editing
\\
\textbf{MN}: Methodology; Supervision; Writing – Review \& Editing
\\
\textbf{AB}: Writing – Review \& Editing
\\
\textbf{MvS}: Writing – Review \& Editing
\\
\textbf{AdRC}: Conceptualization; Methodology; Project Administration; Supervision; Writing – Review \& Editing
\\
\textbf{LMH}: Conceptualization; Methodology; Project Administration; Supervision; Visualization, Writing – Original Draft; Writing – Review \& Editing

\section*{Conflicts of interest}
There are no conflicts to declare.

\section*{Data availability}

All data and code is openly available in the accompanying GitHub repository (\url{https://github.com/OptiMaL-PSE-Lab/mf-bo4bio}) or on Zenodo (\url{https://doi.org/10.5281/zenodo.22659146}).

\section*{Acknowledgements}
The authors wish to thank Prof.~Alexander Gr\"unberger for his advice on modeling uncertainty and differences between experimental scales of bioprocess experimentation.

\newpage
\appendix
\section{Supplementary Information}

\subsection{Clone Parameter Range}
\label{sec:clone_parameters}
\sisetup{
    table-number-alignment=center,
    table-figures-integer=3,
    table-figures-decimal=3,
    separate-uncertainty=true
}
\DeclareSIUnit{\cell}{cell}

\begin{table}[h]
\centering
\caption{\textbf{Lower and upper bounds for process parameter generation in the simulation study} \\ The subscripts X, G, L, A, Q denote viable cells, glucose, lactate, ammonia and glutamine, respectively.}
\label{tab:process_parameters_bounds}
\begin{tabular}{
    l
    S[table-format=1.3]
    S[table-format=1.3]
}
\hline
\textbf{Parameter} & \textbf{Lower bound} & \textbf{Upper bound} \\
\hline
$\mu_{\text{max}}$ [\si{\per\hour}] & 0.02 & 0.06 \\
$K_{\text{lysis}}$ [\si{\per\hour}] & 0.03 & 0.05 \\
$k_{\text{d,Q}}$ [\si{\per\hour}] & 0.001 & 0.0015 \\
$k_{\text{d,max}}$ [\si{\per\hour}] & 0.01 & 0.015 \\
$k_{\mu}$ [\si{\per\hour}] & 0.01 & 0.015 \\
$K_\text{L}$ [\si{\text{mM}}] & 130 & 170 \\
$K_\text{A}$ [\si{\text{mM}}] & 35 & 40 \\
$K_\text{G}$ [\si{\text{mM}}] & 0.9 & 1.6 \\
$K_\text{Q}$ [\si{\text{mM}}] & 0.21 & 0.3 \\
$Y_{\text{X,G}}$ [\si{\text{cells}\per\milli\mole}] & 1e7 & 3e8 \\
$Y_{\text{X,Q}}$ [\si{\text{cells}\per\milli\mole}] & 1e8 & 1e10 \\
$Y_{\text{L,G}}$ [–] & 1 & 3 \\
$Y_{\text{A,Q}}$ [–] & 0.6 & 1.0 \\
$Y_{\text{P,X}}$ [\si{\milli\mole\per\text{cell}}] & 1e-8 & 2e-7 \\
$\dot{Y}_{\text{X,G}}$ [\si{\milli\mole\per\text{cell}}] & 3e-6 & 8e-6 \\
$m_\text{G}$ [\si{\milli\mole\per\text{cell}\per\hour}] & 1e-13 & 5e-10 \\
$m_\text{Q}$ [\si{\milli\mole\per\text{cell}\per\hour}] & 3e-12 & 5e-12 \\
$A$ [–] & 2 & 5 \\
$\mathrm{pH}_{\text{opt}}$ & 6.0 & 8.0 \\
$T_{\text{opt}}$ [\si{\celsius}] & 30 & 40 \\
\hline
\end{tabular}
\end{table}

\subsection{Cost Estimation}
\label{sec:cost_estimation}
\begin{table}[ht]
    \centering
    \caption{\textbf{Assumed parameter values for experiment cost calculation}}
    \label{tab:cost_assumptions}
    \resizebox{0.7\textwidth}{!}{
    \begin{tabular}{ccccc}
        \hline
        Variable & Description & \ac{mtp} & \ac{mbr} & Pilot \\
        \hline
        $c_{\text{medium}}$ (€/L) & Cost of experimentation medium & 100 & 100 & 100 \\
        $V_f$ (L) & Reactor volume & 0.001 & 0.25 & 15 \\
        $c_{\text{hr}}$ (€/h) & Personnel cost per hour & 200 & 200 & 200 \\
        $t_f$ (h) & Time per experiment & 0.03 & 0.25 & 3 \\
        $c_{\text{reactor}}$ (€/reactor) & Cost of single-use reactor & 3 & 500 & - \\
        \hline
    \end{tabular}
    }
\end{table}

To estimate the order of magnitude of these costs, the cost model was divided into personnel and material costs, with energy consumption assumed to be negligible in comparison. The cost estimation is performed per single experiment (e.\,g., for \ac{mtp}, this refers to the cost of one well). The assumptions made for this estimatate are summarized in \Cref{tab:cost_assumptions}. 

It is important to note that the preparation time required per industrial-scale batch is assumed to be three hours across all scales. Consequently, the time per individual experiment is calculated by dividing this batch duration by the respective batch size (96 for \ac{mtp}, 12 for \ac{mbr}, and 1 for the pilot reactor).

Based on these assumptions, the estimated cost per experiment can be calculated as:
\begin{equation*}
    c_f = c_{\text{medium}} \cdot V_f + c_{\text{hr}} \cdot t_f + c_{\text{reactor}},
\end{equation*}
where $c_f$ denotes the total cost per experiment, $c_{\text{medium}}$ the cost per liter of growth medium (i.e., a broth containing nutrients), $V_f$ the reactor volume, $c_{\text{hr}}$ the hourly labor cost, $t_f$ the time per experiment, and $c_{\text{reactor}}$ the cost of a single-use reactor (if applicable).

\subsection{Higher Embedding Dimensions}
\label{sec:higher_dimensions}
The tests conducted with five, six and ten-dimensional entity embeddings, which can be seen in \Cref{tab:clone_representation_dist_a}, show only little performance improvement over the three-dimensional one. Further increasing the dimensionality of the embedding space past a certain point seems to reduce the performance of the presented framework. Similar results have previously been obtained by Hutter \textit{et al.}~\cite{entity_embedding}.

\begin{table}[h]
  \centering
  \caption{\textbf{Comparison of different clone representation strategies}\\
  The table presents the final cumulative maximum endpoint concentrations (mean and standard deviation) for three acquisition functions and varying clone representations across 10 runs for Distribution A. The highest performing representation for each acquisition function is highlighted in bold.}
  \label{tab:clone_representation_dist_a}
\begin{tabular*}{0.52\textwidth}{@{}l@{\hspace{3pt}}l@{\hspace{3pt}}c@{\hspace{3pt}}c@{}}
    \hline
    Method & Clone Representation & Mean (mg/L) & Std (mg/L) \\
    \hline
    \ac{gibbon} & \ac{icm} & 54 & $\pm$ 15 \\
     & Entity Embedding (2D) & 65 & $\pm$ 14 \\
     & \textbf{Entity Embedding (3D)} & \textbf{66} & \textbf{$\pm$ 12} \\
     & Entity Embedding (5D) & 52 & $\pm$ 14 \\
     & Entity Embedding (6D) & 58 & $\pm$ 14 \\
     & Entity Embedding (10D) & 60 & $\pm$ 17 \\
    \ac{qlogei} & \textbf{\ac{icm}} & \textbf{58} & \textbf{$\pm$ 12} \\
     & Entity Embedding (2D) & 51 & $\pm$ 16 \\
     & Entity Embedding (3D) & 54 & $\pm$ 11 \\
     & Entity Embedding (5D) & 49 & $\pm$ 19 \\
     & Entity Embedding (6D) & 46 & $\pm$ 20 \\
     & Entity Embedding (10D) & 46 & $\pm$ 16 \\
    \ac{qucb} & \ac{icm} & 53 & $\pm$ 11 \\
     & \textbf{Entity Embedding (2D)} & \textbf{61} & \textbf{$\pm$ 12} \\
     & Entity Embedding (3D) & 56 & $\pm$ 18 \\
     & Entity Embedding (5D) & 57 & $\pm$ 14 \\
     & Entity Embedding (6D) & 47 & $\pm$ 14 \\
     & Entity Embedding (10D) & 51 & $\pm$ 14 \\
    \hline
\end{tabular*}
\end{table}

\subsection{Fidelity Level Evaluation Distribution Plots}
\label{sec:fid_level_dist}
\begin{figure}[ht]
    \centering
    \includegraphics[width=1\linewidth]{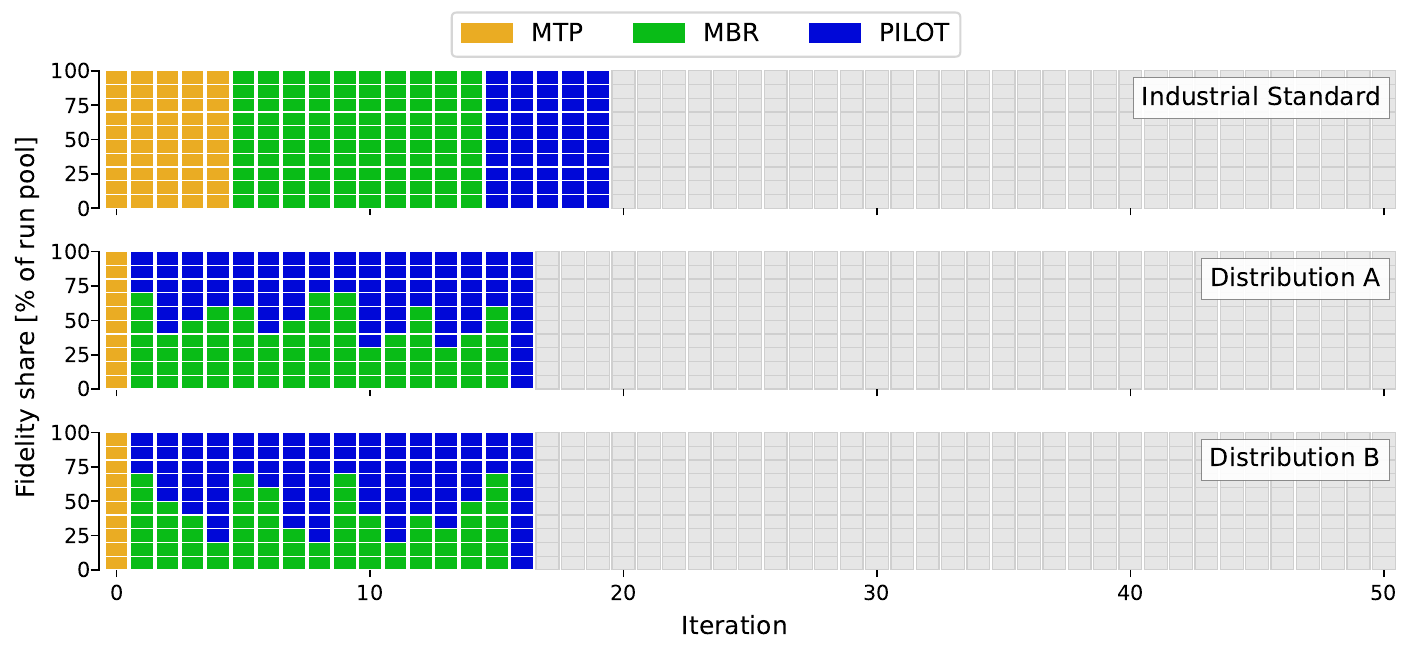}
    \caption{\textbf{Overview of used fidelity levels throughout BO iterations with \ac{gibbon}}\\
    The plot shows the fraction of \ac{gibbon} evaluations at each fidelity level per iteration for Distributions~A and B. The first iteration consists solely of \ac{mtp} evaluations (yellow), which can be followed by a mixture of scales with \ac{mbr} (green) and pilot experiments (blue). The iterations for this acquisition function are dominated by a mixture of \ac{mbr} evaluations with mainly pilot scale experiments.}
    \label{fig:fidelity_distribution_gibbon}
\end{figure}

\begin{figure}[ht]
    \centering
    \includegraphics[width=1\linewidth]{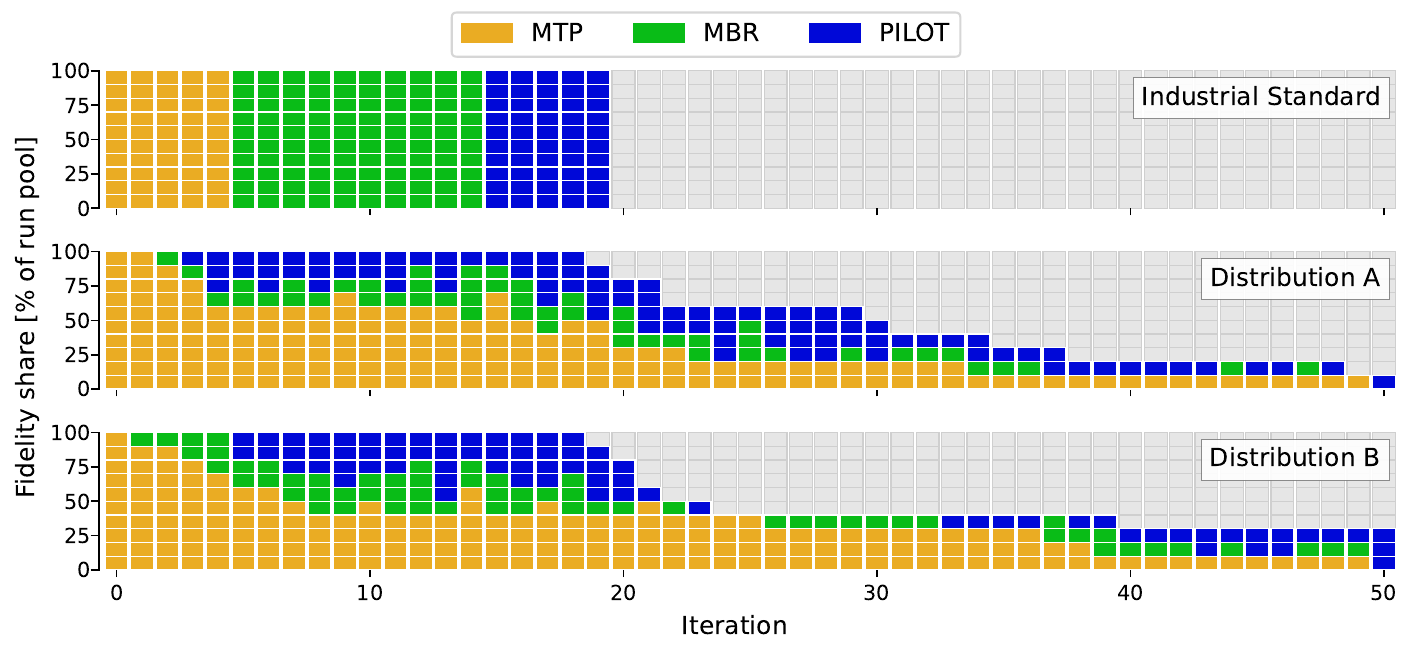}
    \caption{\textbf{Overview of used fidelity levels throughout BO iterations with \ac{qucb}}\\
    The plot shows the fraction of \ac{qucb} evaluations at each fidelity level per iteration for Distributions A and B. The first iteration consists solely of \ac{mtp} evaluations (yellow), which can be followed by a mixture of scales with \ac{mbr} (green) and pilot experiments (blue). The pattern for this acquisition function is characterized primarily by \ac{mtp} evaluations, with a gradual transition to runs at higher fidelity levels during the middle iterations. This tendency is more pronounced for Distribution~A, though still far less so than observed with \ac{qlogei}.}
    \label{fig:fidelity_distribution_qucb}
\end{figure}

Similar patterns to those in \Cref{fig:fidelity_distribution_qlogei} for \ac{qlogei} are observed in \Cref{fig:fidelity_distribution_qucb} for \ac{qucb}. However, \Cref{fig:fidelity_distribution_gibbon} shows that the implemented \ac{gibbon} acquisition function has a pronounced tendency toward pilot-scale evaluations as discussed in \Cref{sec:results_and_discussion}. Interestingly a periodic increased and decreased selection of pilot-scale experiments is visible, where over 3 to 4 iterations the fraction of pilot evaluations increases before dropping off again. For \Cref{fig:fidelity_distribution_qucb} an increase in evaluations on higher fidelity levels is recorded (between iterations 20 and 30). It is also visible through the shorter runtime caused by more frequent use of expensive experiments. Here one might hypothesize that this is due to the lack of feeding information in \ac{mtp} experiments.

\subsection{Effect of a Flexible Feeding Pulse in MTP Experiments}
\label{sec:flexible_feeding}

\begin{figure*}[ht!]
    \centering
    \includegraphics[width=1\linewidth]{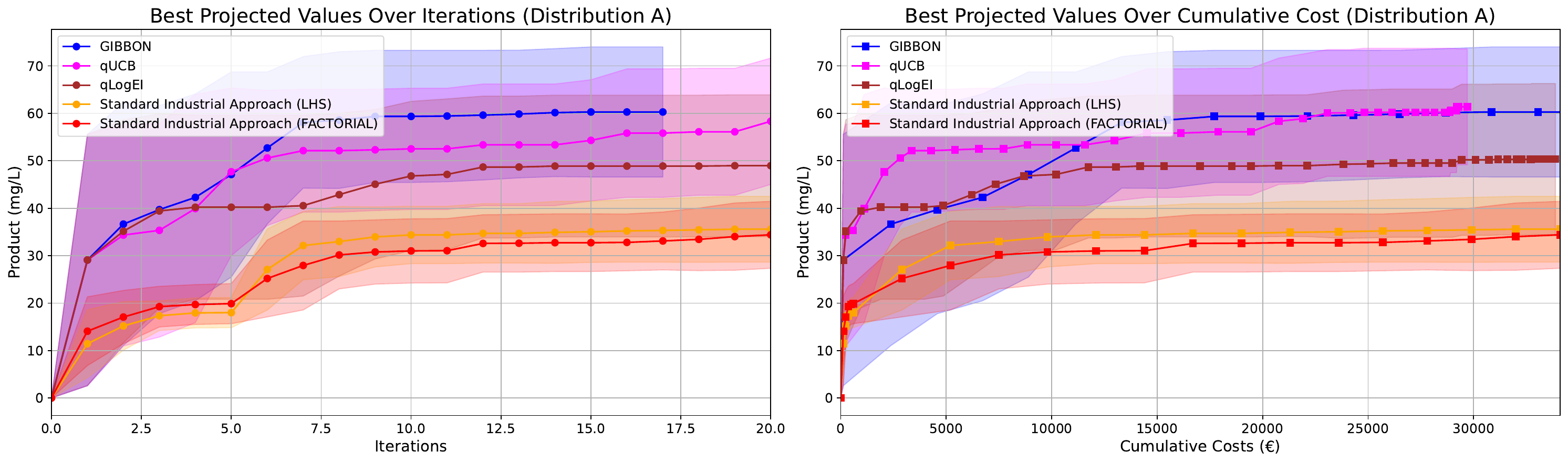}
    \caption{\textbf{Comparison of BO results for Distribution~A with Flexible Feeding in \ac{mtp}-scale} \\
    Showing the best found projected values for multiple optimization approaches. This plot compares the performance of \ac{gibbon} (blue), \ac{qlogei} (brown), \ac{qucb} (magenta), the industrial approaches with factorial design (red) and Latin Hypercube sampling (yellow). For this scenario a distribution of clones with more misaligned scale-behavior was used (see Distribution~A in \Cref{fig:clone_distribution}). On the left side the behavior is plotted over the first 20 iterations while on the right it is shown over cumulative cost of evaluation up to €34,100. The plots enable a comparison of the optimization behavior within the iteration and cost boundaries of the industrial benchmark.}
    \label{fig:results_dist_A_flexible_feeding}
\end{figure*}

When introducing flexible feeding (removing the constraint $F_\text{max}=0$), a clear contrast emerges between the two data distributions regarding how traditional industrial methods, such as \ac{lhs} and Factorial Design, handle clone selection and scale transitions compared to \ac{bo}. In Distribution~A, the benefits of flexible feeding are highly pronounced for the \ac{bo} methods as can be seen in \Cref{fig:results_dist_A_flexible_feeding}. Specifically, \ac{qucb} achieves a peak concentration of \SI{61}{\milli\gram\per\litre}, representing a \SI{70}{\percent} improvement over the best industrial baseline while simultaneously reducing total costs by \SI{13}{\percent}. Conversely, the industrial methods struggle to leverage the flexible feeding conditions, showing improvement only during the first five iterations and stalling at a peak of \SI{36}{\milli\gram\per\litre}, which is remarkably similar to the results seen in the non-feeding scenario. This early plateau could occur because the industrial methods fixate early on well-performing clones on \ac{mtp}-scale, specifically clones 14, 15, and 29 (see \Cref{fig:clone_distribution} in Distribution~A). This early convergence signifies a detrimental lock-in effect during scale transitions, preventing the discovery of potentially superior clone-condition combinations and allowing \ac{bo} methods to easily surpass the best industrial peak. 

\begin{figure*}
    \centering
    \includegraphics[width=1\linewidth]{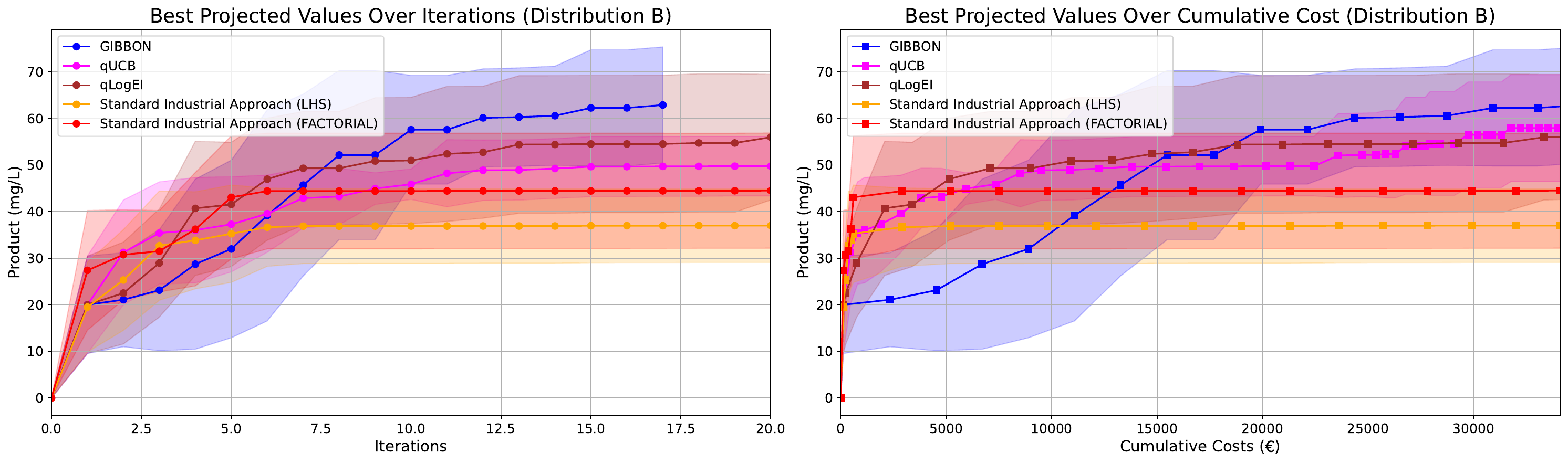}
    \caption{\textbf{Comparison of BO results for Distribution~B with Flexible Feeding in \ac{mtp}-scale} \\
    Showing the best found projected values for multiple optimization approaches. This plot compares the performance of \ac{gibbon} (blue), \ac{qlogei} (brown), \ac{qucb} (magenta), with the industrial approaches \ac{lhs} (yellow) and factorial design (red). For this scenario a distribution of clones with higher behavior across scales was used (see Distribution~B in \Cref{fig:clone_distribution}). On the left side the behavior is plotted over the first 20 iterations while on the right it is shown over cumulative cost of evaluation up to €34,100. The plots enable a comparison of the optimization behavior within the iteration and cost boundaries of the industrial benchmark.}
    \label{fig:results_dist_B_flexible_feeding}
\end{figure*}

Distribution~B, however, demonstrates a landscape better suited for early exploration. Here, the industrial methods perform much better, with the Factorial approach reaching \SI{45}{\milli\gram\per\litre}, visible in \Cref{fig:results_dist_B_flexible_feeding}. Because the initial clone selection is much more informative, the traditional funnel-like approach keeps pace with \ac{bo} during the early stages, ensuring that better clones are paired with optimized conditions in later iterations. Despite this competitive early performance, \ac{bo} methods still dominate the final results, with \ac{gibbon} hitting \SI{63}{\milli\gram\per\litre}, a \SI{41}{\percent} improvement over the Factorial method. While the \ac{bo} methods incur a marginal total cost increase of approximately \SI{3}{\percent} compared to industrial baselines, their ability to push the absolute peak concentration makes them highly cost-effective in terms of yield. Ultimately, these results highlight a crucial dynamic in bioprocess optimization. While advanced flexible feeding setups with well-aligned \ac{mtp}-scale experiments and classical methods can yield promising results, \ac{bo} with mixed-fidelity experimentation remains the superior and more robust choice. It drives better global exploration and actively prevents the scale-up lock-in effect observed in Distribution~A, making it the recommended approach for maximizing result generation regardless of the data landscape.

\clearpage
%Bibliography
\bibliographystyle{unsrt}  
\bibliography{references}

\end{document}